\documentclass[structabstract]{aa}  

\usepackage{graphicx}
\usepackage{natbib}
\usepackage{bm}
\usepackage{txfonts}
\begin{document}
   \title{On Self-Sustained Dynamo Cycles in Accretion Discs}


   \author{G. Lesur
          \inst{1}
          \and
          G. I. Ogilvie\inst{1}
          }
   \institute{Department of Applied Mathematics and Theoretical Physics, University of Cambridge, Centre for Mathematical Sciences,
Wilberforce Road, Cambridge CB3 0WA, UK \\
              \email{g.lesur@damtp.cam.ac.uk}
                     }

   \date{Received 7 May 2008 / Accepted 6 July 2008}

 
  \abstract
   {MHD turbulence is known to exist in shearing boxes with either zero or nonzero net magnetic flux. However, the way turbulence survives in the zero-net-flux case is not explained by linear theory and appears as a purely numerical result that is not well understood. This type of turbulence is also related to the possibility of having a dynamo action in accretion discs, which may help to generate the large-scale magnetic field required by ejection processes.}
   {We look for a nonlinear mechanism able to explain the persistence of MHD turbulence in shearing boxes with zero net magnetic flux, and potentially leading to large-scale dynamo action.}
   {Spectral nonlinear simulations of the magnetorotational instability are shown to exhibit a large-scale axisymmetric magnetic field, maintained for a few orbits. The generation process of this field
   is investigated using the results of the simulations and an inhomogeneous linear approach. We show that quasilinear nonaxisymmetric waves may provide a positive back-reaction on the large-scale field when a weak inhomogeneous azimuthal field is present, explaining the behaviour of the simulations. We finally reproduce the dynamo cycles using a simple closure model summarising our linear results.}
   {The mechanism by which turbulence is sustained in zero-net-flux shearing boxes is shown to be related to the existence of a large-scale azimuthal field, surviving for several orbits. In particular, it is shown that MHD turbulence in shearing boxes can be seen as a dynamo process coupled to a magnetorotational-type instability.}
   {}

   \keywords{accretion, accretion disks -- MHD -- turbulence}

   \maketitle
%

\section{Introduction}

  The problem of angular momentum transport is a central issue of accretion disc theory. Following \cite{SS73}, angular momentum transport is often modelled assuming the disc is turbulent, using a kind of turbulent viscosity (the so-called $\alpha$ model). However, the way discs may become turbulent is still a highly debated subject. A possible route to turbulence is the magnetorotational instability \citep[MRI,][]{BH91a} which appears in discs sufficiently ionised to be coupled with the magnetic field lines \citep{G96}. This instability has been extensively studied in its nonlinear regime using local \citep{HGB95,SHGB96} and global \citep{H00} numerical simulations. More recently, the effect of non-ideal MHD has been investigated numerically in shearing boxes
with either zero \citep{FPLH07} or nonzero \citep{FSH00,LL07} net magnetic flux, showing a strong dependence on the magnetic Prandtl number $Pm$. 
  
  These new results bring back the question of the efficiency of MHD turbulence in accretion discs since $Pm$ is believed to vary by several orders of magnitude in real objects \citep{BH08}. In particular, as turbulence seems to disappear in zero-net-flux simulations for $Pm\le 1$, the existence of a turbulent flow in low-$Pm$ objects such as protoplanetary discs is questionable. However, as pointed out by \cite{FPLH07}, today simulations can reach Reynolds numbers that are very small (up to $10^4$) compared to those of real discs ($\sim 10^{10}$). Therefore, no clear conclusion for accretion discs can be drawn from these numerical results. 
  
  It has been pointed out by \cite{PCP07} and \cite{FP07} that numerical resolution can also play an important role in simulations. In particular, increasing resolution in zero-net-flux simulations \emph{without} any physical dissipation leads to a weaker turbulence and transport. One might conclude from these results that turbulence should disappear in real astrophysical systems \citep{PCP07}. We think however that this conclusion is questionable since the ideal MHD model does not hold in turbulent flows, mainly because a dissipation scale necessarily exists, at which non-ideal effects are \emph{not} negligible. Of course, ideal MHD simulations must include some sort of numerical dissipation in their algorithm, which then implicitly defines a dissipation scale of the order of the numerical grid scale. However, this algorithm-dependent dissipation is quite different from physical dissipation processes \citep{LB07}, leading potentially to numerical artifacts for large-scale properties, such as turbulent transport.
       
On the other hand, the way MHD turbulence is sustained in these simulations is not well understood. It is known that when a mean vertical field is applied, the magnetorotational instability destabilises the flow and leads to developed three-dimensional turbulence \citep{BH91a,HGB95}. This picture is not directly applicable to zero-net-flux simulations since the magnetic field can be dissipated either by a finite resistivity, or by the turbulence itself (see \cite{HGB96} for an extensive discussion of this point). Therefore, even if one assumes that the MRI appears locally because of a given magnetic field configuration, one has to regenerate this field with some sort of dynamo mechanism. The whole process sustaining turbulence in zero-net-flux simulations is therefore a \emph{nonlinear} mechanism, potentially involving a kind of magnetorotational instability at some stage. As already mentioned by \cite{BH92}, this means that this disc dynamo is intrinsically nonlinear, as it requires a non-negligible Lorentz force, and cannot be described as a kinematic dynamo. Following these ideas, \cite{ROP07} obtained a steady nonlinear solution to the MHD equations in the zero-net-flux case, using rigid conducting walls as radial boundary conditions. This solution is clearly a first step toward the understanding of the mechanism working in shearing boxes, although the boundary conditions and Reynolds numbers are significantly different compared to turbulent simulations. Note also that \cite{BNST95} studied this dynamo process in discs, using boundary conditions allowing for mean flux variations. Although a azimuthal field was generated in their simulations, no physical understanding of the underlying process was provided.

In this paper, we describe a possible mechanism able to sustain MHD turbulence in the shearing box with zero net flux. First, we recall the resistive MHD equations in the shearing box, and the numerical method used to solve them. Next, we investigate the temporal evolution of some zero-net-flux simulations, and we exhibit a long-timescale cycle for the large-scale magnetic field. The origin of this cycle is described using a spectral decomposition, and we demonstrate that its origin is related to some properties of the nonaxisymmetric structures of the flow. We then study a linear theory of nonaxisymmetric waves in the presence of a large-scale magnetic field similar to the one observed in simulations. We show that this linear theory predicts the right properties for the nonaxisymmetric structures and partially explains the nonlinear cycle. We then provide a phenomenological closure model for the evolution of the large-scale field, based on our previous findings. We show that this model reproduces the basic behaviour of the cycles, and provides a mechanism able to sustain turbulence in dissipative MHD shearing boxes. The existence of a long-timescale cycle and large-scale structures in these simulations is the most significant finding of our investigation, mainly because it shows that MHD turbulence is able to generate large scales, independent of the dissipative scales. This mechanism can also be seen as a way to generate large-scale magnetic fields, providing a dynamo mechanism specific to accretion disc turbulence. These findings are discussed in detail in the concluding section, along with the implications for real astrophysical systems.

 \begin{figure*}
   \centering
   \includegraphics[width=1.0\linewidth]{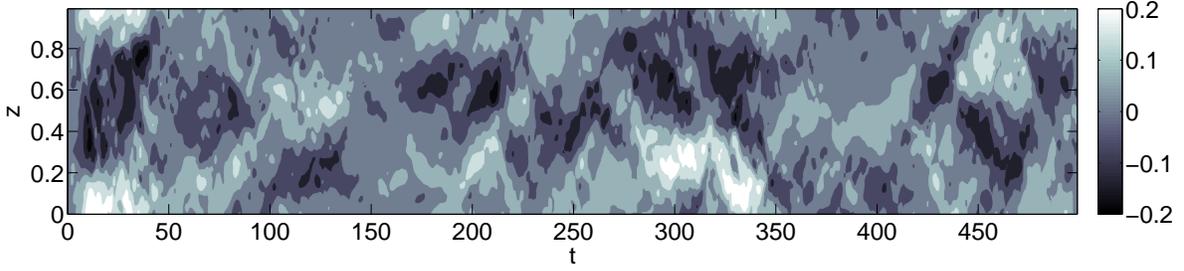}
   \caption{Time-evolution of $B_x$, averaged in the $x$ and $y$ directions. Notice the long-lived ($T\sim 50\, S^{-1}$) vertical structures, mostly found with long vertical wavelengths.}
              \label{bxtm}%
    \end{figure*}

\section{Shearing-box equations and numerical method}
MRI-related turbulence has been extensively studied in the literature. Therefore, we will recall here briefly the basic equations for the shearing-box model. The reader may consult \cite{HGB95}, \cite{B03} and \cite{RU08} for an extensive discussion of the properties and limitations of this model. Since MHD turbulence in discs is subsonic, we will work in the incompressible approximation, which allows us to eliminate sound waves or density waves. We also neglect vertical stratification, consistent with the local shearing-box model \citep{RU08}.
We include in our description a molecular viscosity and resistivity to minimize the artifacts of numerical dissipation.

The shearing-box equations are found by considering a Cartesian box centred at $r=R_0$, rotating with the disc at angular velocity $\Omega=\Omega(R_0)$ and having dimensions $(L_x,L_y,L_z)$ with $L_i \ll R_0$.
We define $R_0\phi \rightarrow x$ and $r-R_0 \rightarrow -y$ for consistency with the standard notation for plane Couette flow \citep[e.g. ][]{DR81}. Note that this definition differs from the standard notation used in shearing boxes \citep{HGB95} with $x\rightarrow -y_\mathrm{SB}$, $y\rightarrow x_\mathrm{SB}$ and $z\rightarrow z_\mathrm{SB}$. In this rotating frame, one eventually obtains the following set of equations:
\begin{eqnarray}
\nonumber \partial_t \bm{u}+\bm{\nabla\cdot} (\bm{u\otimes u})&=&-\bm{\nabla} \Pi+\bm{\nabla\cdot }(\bm{ B \otimes B}) \\
& &-2\bm{\Omega \times u}+2\Omega S y \bm{e_y}+\nu \bm{\Delta u},\\
\partial_t \bm{B}&=&\bm{\nabla \times} (\bm{u \times B}) +\eta \bm{\Delta B},\\
\label{divv} \bm{\nabla \cdot u}&=&0,\\
\bm{\nabla \cdot B}&=&0.
\end{eqnarray}
The boundary conditions associated with this system are periodic in the $x$ and $z$ direction and shearing-periodic in the $y$ direction (see \cite{HGB95} for a complete description of these boundary conditions). 
In these equations, we have defined the mean shear $S=-r\partial_r \Omega$, which is set to $S=(3/2)\Omega$, assuming a Keplerian rotation profile. The generalized pressure term $\Pi$ includes both the kinematic pressure $P/\rho_0$ and the magnetic pressure $\bm{B}^2/2$. One should note that the generalized pressure $\Pi$ is actually a Lagrange multiplier enforcing equation (\ref{divv}), and is therefore computed solving a Poisson equation. Note also that the magnetic field is expressed in Alfv\' en-speed units, for simplicity. 

The steady-state solution to these equations is the local Keplerian profile $\bm{u}=Sy\bm{e_x}$. In this paper, we will consider only the turbulent deviations from this Keplerian profile.  These may be written as $\bm{v}=\bm{u}-Sy\bm{e_x}$, leading to the following equations for $\bm{v}$:

\begin{eqnarray}
\nonumber \partial _t \bm{v}+\bm{\nabla \cdot}(\bm{v \otimes v})&=&-\bm{\nabla} \Pi+\bm{\nabla\cdot }(\bm{ B \otimes B})-Sy\partial_x \bm{v}\\
\label{motion}& & +(2\Omega-S) v_y\bm{e_x}-2\Omega v_x \bm{e_y}+\nu\bm{\Delta v},\\
 \nonumber \partial _t \bm{B}&=&-Sy\partial_x \bm{B}+SB_y\bm{e_x}\\
 \label{induction}& & +\bm{\nabla \times} (\bm{v \times B}) +\eta \bm{\Delta B},\\
\label{Vstruct} \bm{\nabla \cdot v}&=&0,\\
\label{Bstruct}\bm{\nabla \cdot B}&=&0.
\end{eqnarray}
Following \cite{HGB95}, one can integrate the induction equation (\ref{induction}) over the volume of the box, leading to:
\begin{equation}
\frac{\partial \langle \bm{B}\rangle }{\partial t}=S\langle B_y\rangle \bm{e_x},
\end{equation}
where $\langle \rangle$ denotes a volume average. Therefore, the mean magnetic field is conserved, provided that no mean radial field is present. This allows us to define the zero-net-flux shearing box, as the box in which $\langle \bm{B} \rangle=0$. 

To numerically solve the shearing-box equations, we use a spectral Galerkin representation of equations (\ref{motion})--(\ref{Bstruct}) in the sheared frame \citep[see][]{LL05}. This frame allows us to use a Fourier decomposition since the shearing-sheet boundary conditions are transformed into perfectly periodic boundary conditions. Moreover, this decomposition allows us to conserve magnetic flux to machine precision without any modification, which is an advantage compared to finite-difference or finite-volume methods (the total magnetic flux created during one simulation is typically $10^{-11}$). Equations (\ref{Vstruct}) and (\ref{Bstruct}) are enforced to machine precision using a spectral projection \citep{P02}. The nonlinear terms are computed with a pseudospectral method, and aliasing is prevented using the 3/2 rule. To always compute the physically relevant scales in the sheared frame, we use a remap method similar to the one described by \cite{UR05}. This routine redefines the sheared frame every $T_{\mathrm{remap}}=L_x/(L_yS)$ and we have checked that none of the behaviour we describe in this paper was related to this numerical timescale. Since spectral methods are very little dissipative by nature, we check that numerical dissipation is kept to very small values, computing the total energy budget at each time step \citep[see][for a discussion of this procedure]{LL05}. We therefore ensure that numerical dissipation is responsible for less than $3\%$ of the total dissipative losses occurring in these simulations.

To quantify the dissipation processes in the
simulations, we use dimensionless numbers defined as:
\begin{itemize}
\item The Reynolds number, $Re=SL_z^2/\nu$, comparing the nonlinear advection term to the viscous dissipation.
\item The magnetic Reynolds number, $Rm=SL_z^2/\eta$, comparing magnetic field advection to the Ohmic resistivity.
\item The magnetic Prandtl number, defined as the ratio of the two previous quantities $Pm=Rm/Re=\nu/\eta$, which measures the relative importance of the dissipation processes.
\end{itemize}
In the following we use $S^{-1}$ as the unit of time and $SL_z$ as the unit of velocity.
One orbit corresponds to $T_\mathrm{orb}=3\pi S^{-1}$.

     \begin{figure}
   \centering
   \includegraphics[width=1.0\linewidth]{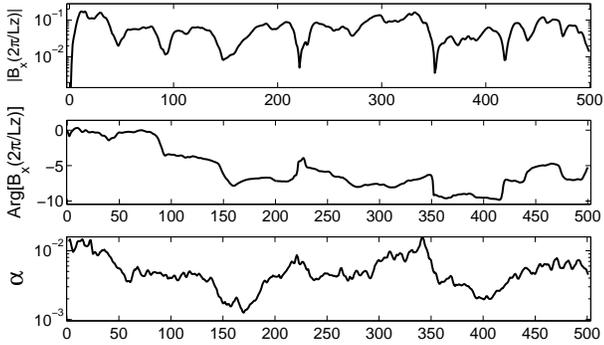}
   \caption{Fourier analysis of $\widehat{B_x}(k_x=0,k_y=0,k_z=2\pi/L_z)$: amplitude (top panel) and phase (middle panel). This mode exhibits long-timescale ($T\sim 50\, S^{-1}$) cycles during which the phase is approximately constant. The transport coefficient ($\alpha$, bottom panel) does not show this cyclic behaviour.}
              \label{SimRe800}%
    \end{figure}
    
\section{Simulations of zero-net-flux MHD turbulence\label{sectioncycle}}
\subsection{Long-timescale cycle in zero-net-flux MHD turbulence}
The first zero-net-flux turbulent flow was computed by \cite{SHGB96} in the context of a stratified compressible shearing box. In this section we consider the simpler unstratified and incompressible case. The aspect ratio is set to $L_y=L_z/2$ and $L_x=2L_z$, which corresponds to the box used by \cite{LL07} elongated twice in the vertical direction. As we will see, a box elongated in the $z$ direction is useful to exhibit the large-scale and long-lived vertical structures one may miss with more classical configurations. The resolution of the simulation presented in this section is $n_x\times n_y\times n_z=128\times 64 \times 128$. This corresponds approximately to an ``equivalent'' resolution of $256\times 128 \times 256$ for a second-order finite-difference method \citep{FPLH07}. 

We present in Fig.~\ref{bxtm} the evolution of the azimuthal component of the magnetic field averaged in the $x$ and $y$ directions as a function of $z$ and $t$, for a simulation with $Re=3200$, $Pm=4$. This component clearly shows some long-lived structures, mostly of large vertical wavelength. This result is at first sight surprising since one expects turbulent structures to be modified on a typical dynamical timescale, i.e.\ $S^{-1}$. To quantify this effect more precisely, we plot in Fig.~\ref{SimRe800} the time history of the largest vertical Fourier mode, defined by:
\begin{equation}
\label{FourierT}\widehat{B_x}(k_z)=\frac{1}{L_xL_yL_z}\int_0^{L_x} \!\!\!\int_0^{L_y} \!\!\! \int_0^{L_z} \!\! B_x(x,y,z)\exp(-ik_z z)\, dx\,d y \, d z,
\end{equation}
The amplitude of the largest vertical mode (with $k_z=2\pi/L_z$) shows clearly the cyclic behaviour already seen in Fig.~\ref{bxtm}, with a typical period $T\sim30-50\,S^{-1}$. The phase analysis shows an approximately constant phase during each cycle, meaning that these structures are not moving vertically.  We have checked that these structures are not a transient phenomenon, integrating one simulation up to $t=2000\, S^{-1}$ without any significant modification.  Note that the resistive timescale for the largest-scale mode $k_z=2\pi/L_z$ is $324\,S^{-1}$ in this case ($Rm=12800$).
For comparison with other simulations, we also compute the turbulent transport (Fig.~\ref{SimRe800}, bottom), measured by the Shakura--Sunyaev-like coefficient
\begin{equation}
\alpha=\frac{\langle B_x B_y \rangle-\langle v_x v_y\rangle}{S^2L_z^2}.
\end{equation}
The turbulent transport obtained in our case is comparable with previous simulations ($\alpha\sim 7\times 10^{-3}$). However, the cyclic behaviour described above is not seen in the transport coefficients for this simulation, which partly explains why it has not been observed in previous works. 

 \begin{figure*}
   \centering
   \includegraphics[width=0.45\linewidth]{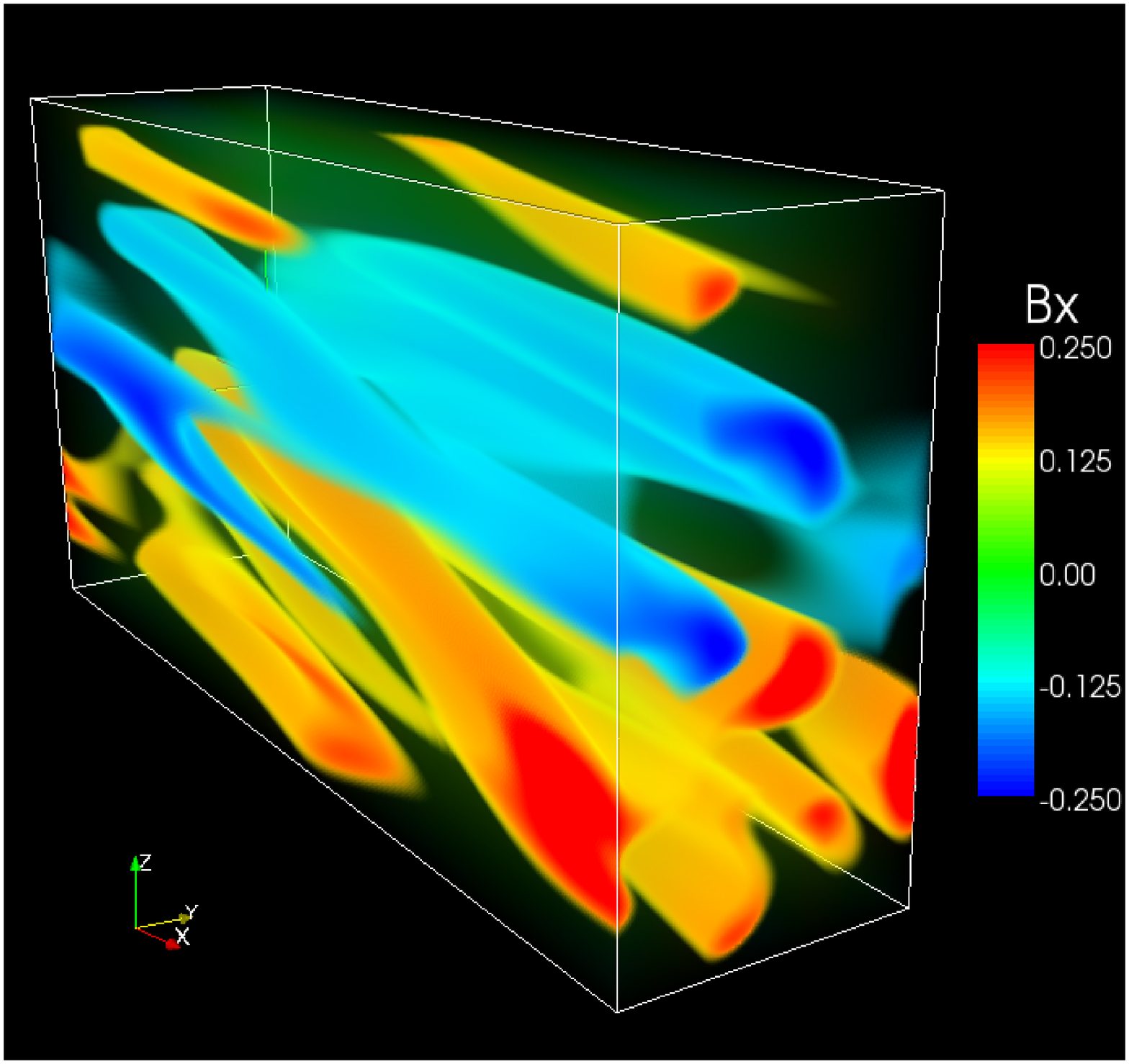}
   \quad
   \includegraphics[width=0.45\linewidth]{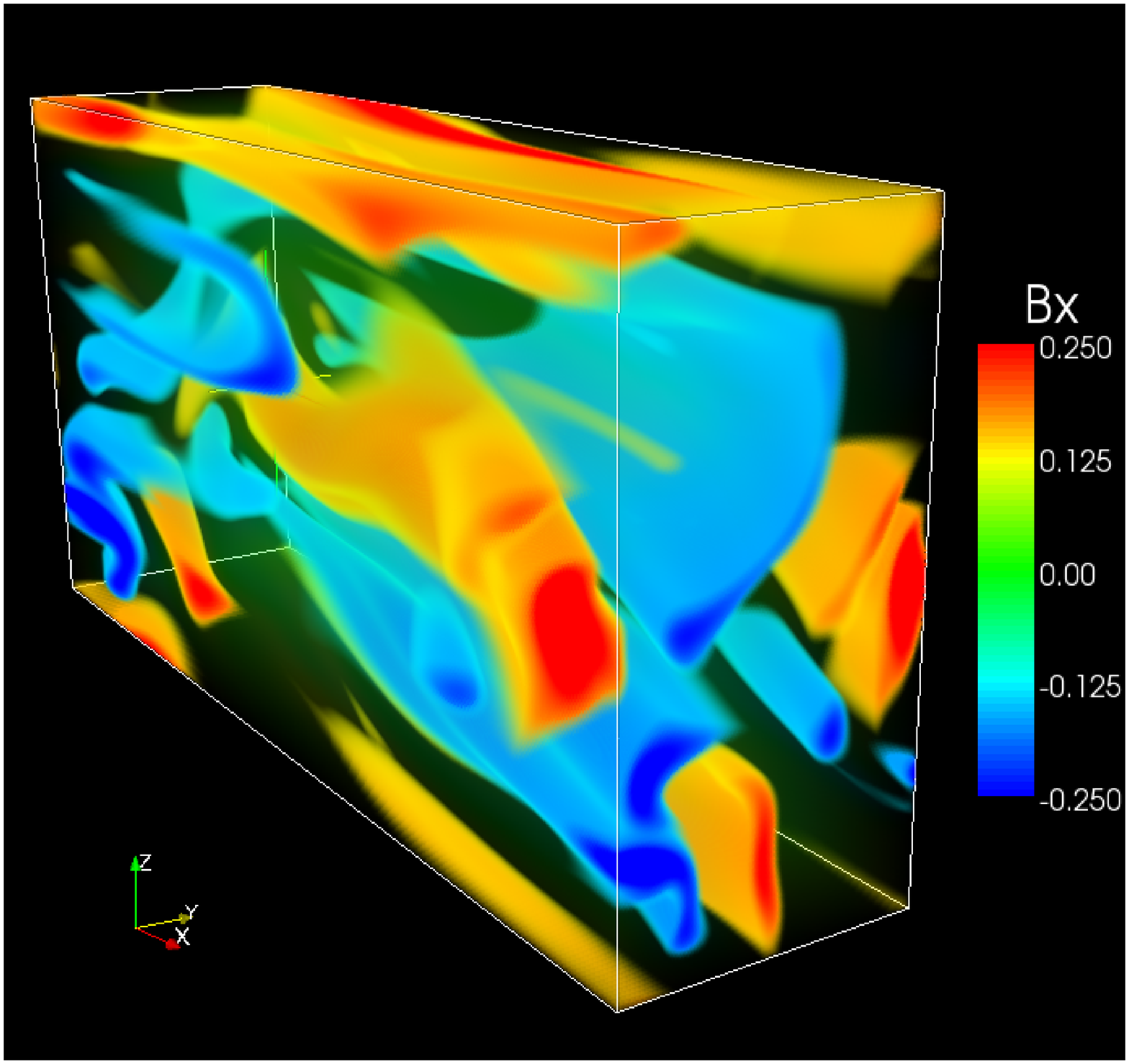}
   \caption{Volume rendering of $B_x$ for a simulation with $Re=1600$ and $Rm=6400$. Left panel: $t=260$, corresponding to a cycle maximum. Right panel: $t=280$, corresponding to a cycle minimum. One easily observes the large-scale $B_x(z)$ on the $t=260$ snapshot whereas strong nonaxisymmetric structures destroy the large scale structures at $t=280$.}
              \label{SimuSnap}%
    \end{figure*}

 \begin{figure*}
   \centering
   \includegraphics[width=0.45\linewidth]{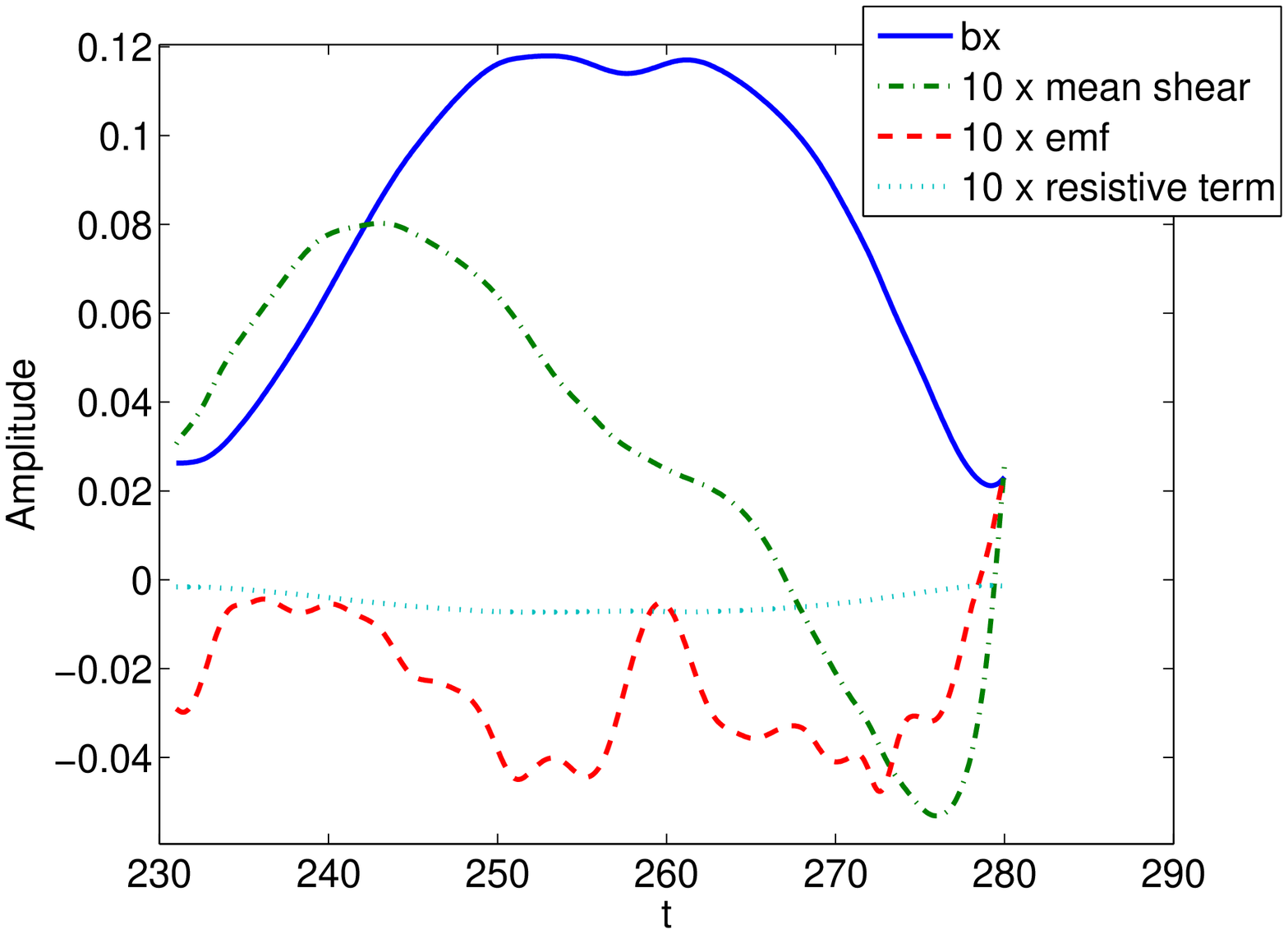}
   \includegraphics[width=0.45\linewidth]{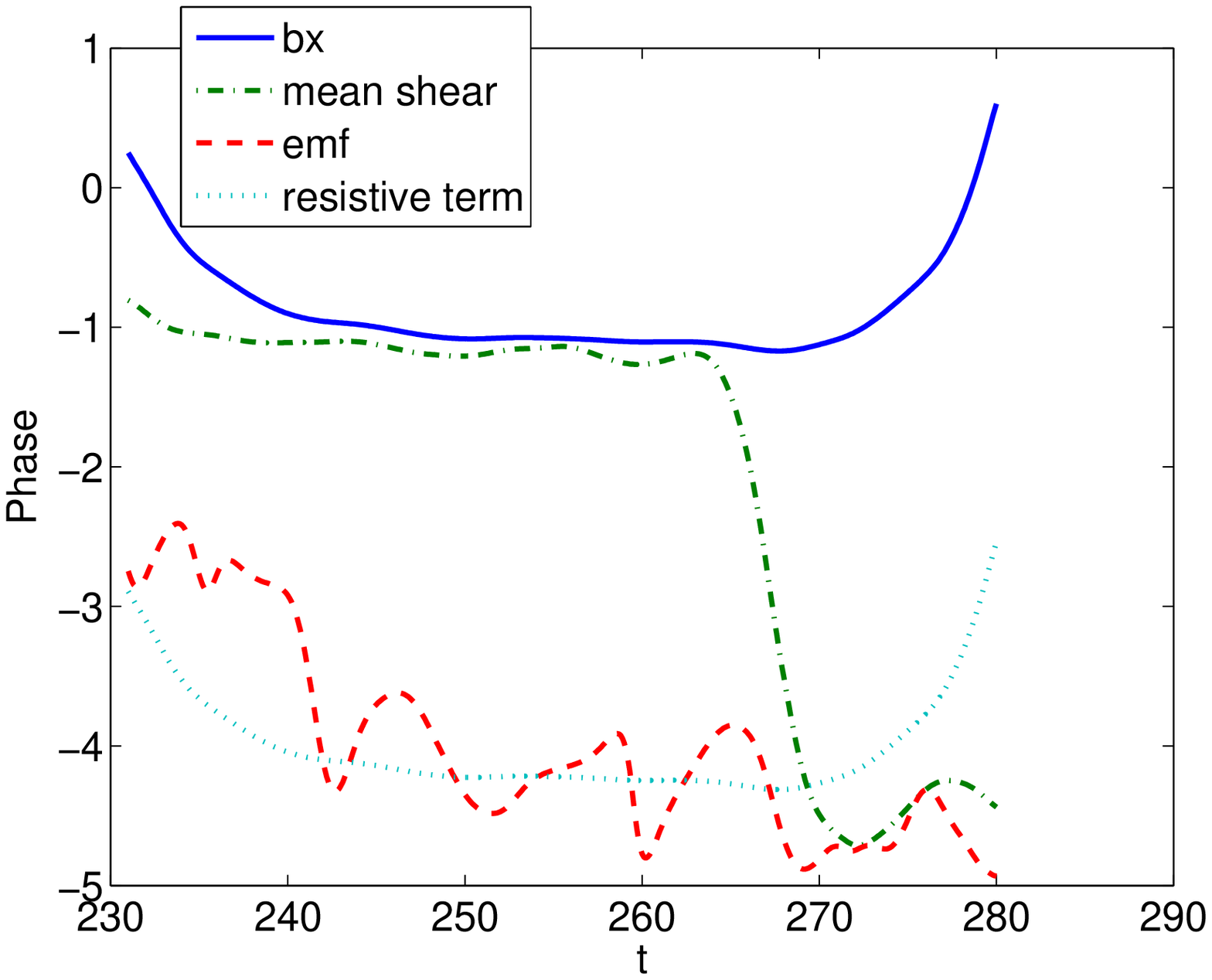}
   \caption{Amplitude projected on $\widehat{B_x}(k_0,t)$ (left) and phase (right) of the terms involved in equation (\ref{bxbudgeteqn}). The amplitude clearly shows one cycle comparable to Fig.~\ref{SimRe800} for $B_x$. The cycle behaviour comes mainly from the shear term, whose contribution is initially positive ($t<270$) and later becomes negative ($t>270$). EMF and resistive term always make dissipative contributions.}
              \label{bxbudget}%
    \end{figure*}
    
    \begin{figure*}
   \centering
   \includegraphics[width=0.45\linewidth]{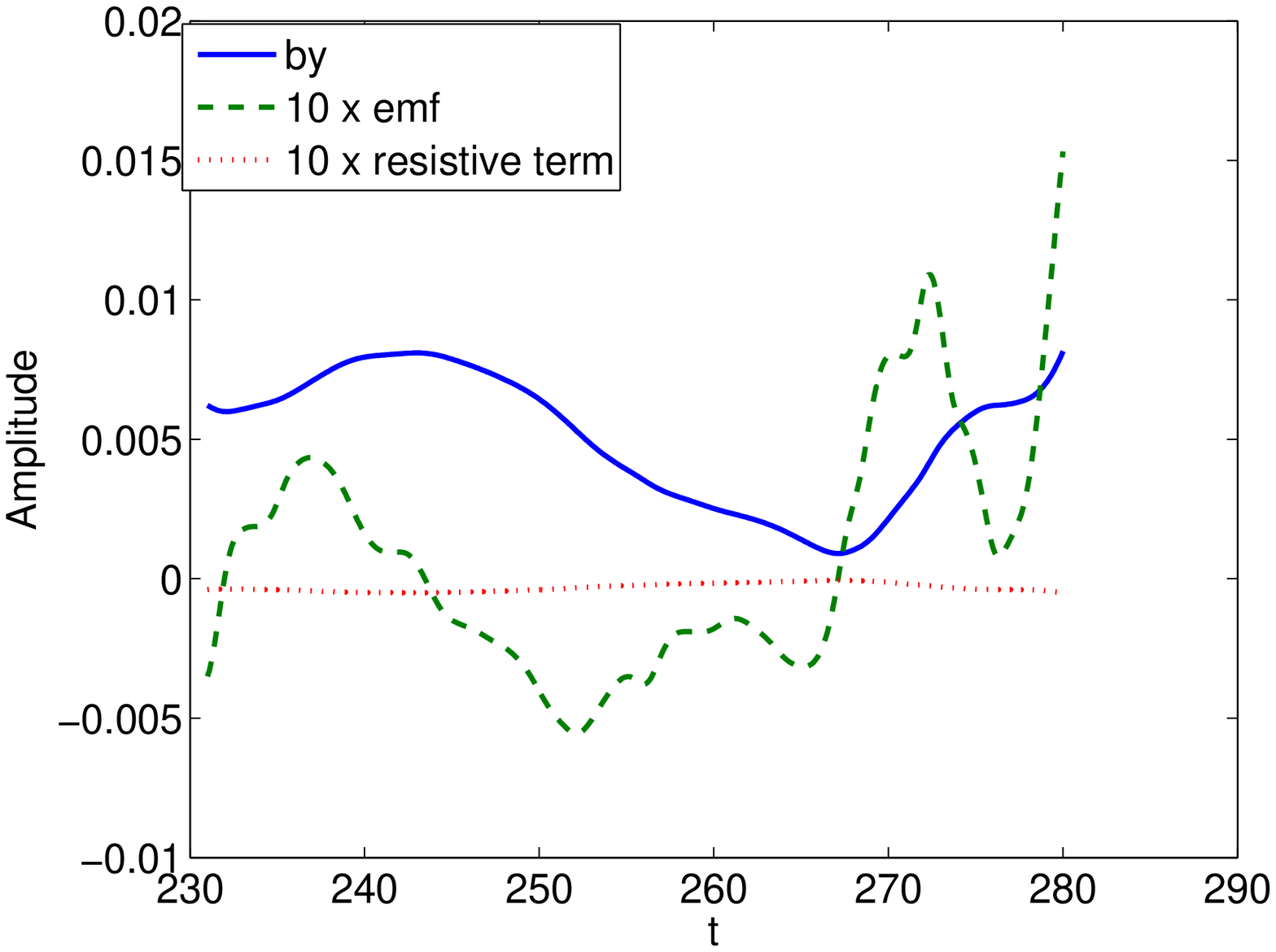}
   \includegraphics[width=0.45\linewidth]{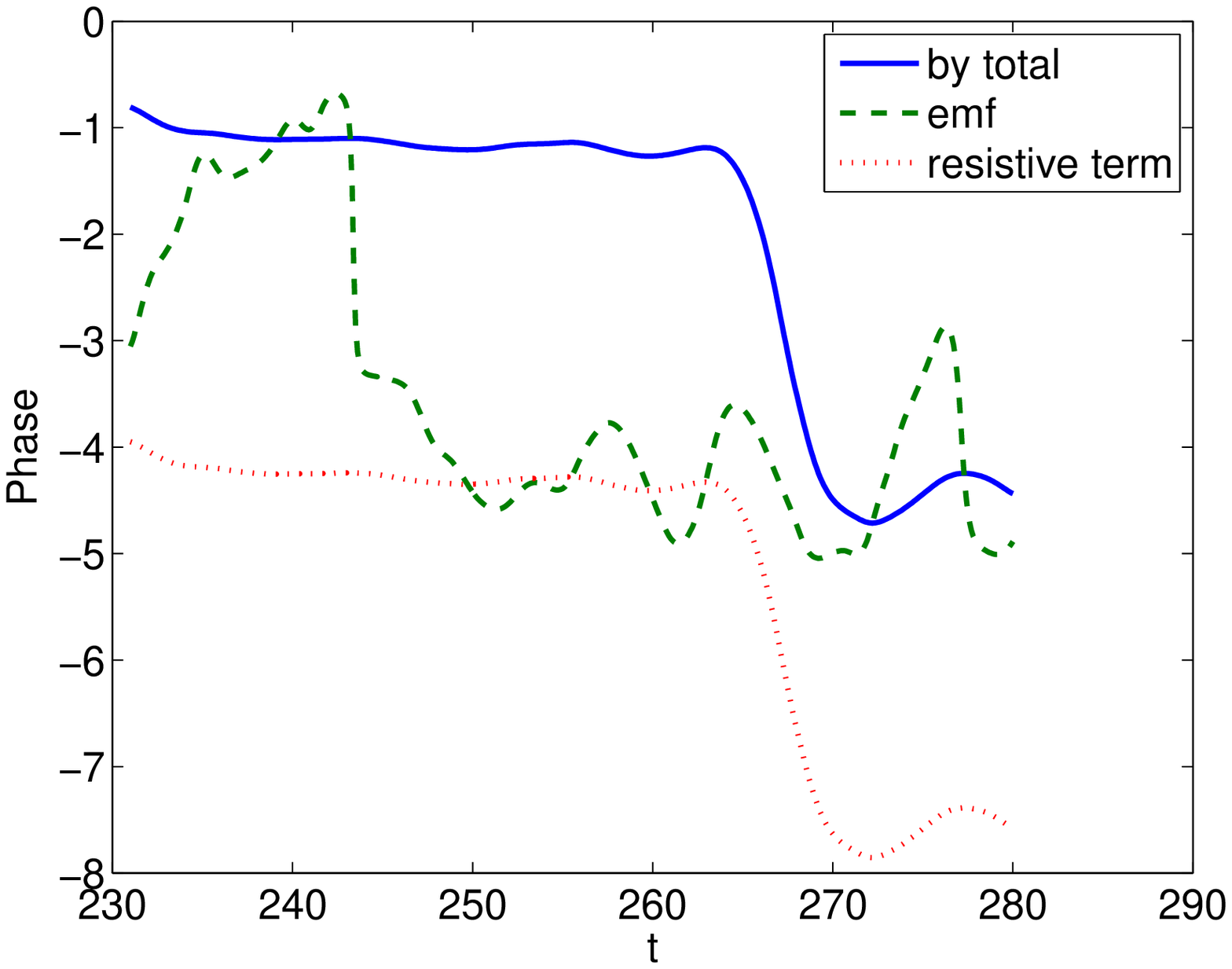}
   \caption{Amplitude projected on $\widehat{B_y}(k_0,t)$ (left) and phase (right) of the terms involved in equation (\ref{bybudgeteqn}). The amplitude of $B_y$ is approximately 20 times smaller than $B_x$. The cycle observed for $B_y$ is related to the oscillation of the EMF $\mathcal{E}_x$ which is reversed at $t\simeq 243$.} 
              \label{bybudget}%
    \end{figure*}

\subsection{Cycle analysis}
Naturally, one may wonder what mechanism generates this magnetic field structure, and whether this mechanism is related to some turbulent transport properties. To investigate these questions, we first reduce the Reynolds number of the simulation, keeping $Pm$ constant. This allows us to have ``cleaner'' flows to work with, removing much of the small-scale variations. We are able to maintain MHD turbulence and cycles down to $Re=1600$, consistent with previous results \citep{FPLH07}. We show in Fig.~\ref{SimuSnap} some snapshots of the azimuthal magnetic field $B_x$, demonstrating the large scale field structure and its destruction, as expected from the Fourier analysis.
We then isolate one cycle from a $Re=1600$, $Pm=4$  simulation and compute the budget of the Fourier component $\widehat{B_x}(k_z=2\pi/L_z)\equiv \widehat{B_x}(k_0)$. Applying the transformation (\ref{FourierT}) to equation (\ref{induction}), this budget may be written as:
\begin{eqnarray}
\label{bxbudgeteqn}
\partial _t \widehat{B_x}(k_0)&=&S\widehat{B_y}(k_0)-ik_0 \widehat{\mathcal{E}_y}(k_0)-\eta k_0^2 \widehat{B_x}(k_0),\\
\label{bybudgeteqn}
\partial _t \widehat{B_y}(k_0)&=&ik_0 \widehat{\mathcal{E}_x}(k_0)-\eta k_0^2 \widehat{B_y}(k_0),
\end{eqnarray}
where we have defined the electromotive force (EMF) $\bm{\mathcal{E}}=\bm{v\times B}$. 
 According to these equations, $\widehat{B_x}(k_0)$ has three sources: the shear of radial magnetic field lines, the radial EMF and the resistivity (acting as a sink) whereas only the azimuthal EMF and resistivity effects appear for $\widehat{B_y}$. 
 To study the contribution of each term to the global behaviour of $(B_x,B_y)$, we project them on the complex vector $(\widehat{B_x},\widehat{B_y})(k_0)$, defining:
\begin{eqnarray}
\label{CorrelationS}
\verb|bj|&=&|\widehat{B_j}(k_0,t)|,\\
\verb|mean shear|&=&\frac{\Re\Big(S\widehat{B_y}(k_0,t)\widehat{B_x}^*(k_0,t)\Big)}{|\widehat{B_x}(k_0,t)|},\\
\label{emfdef}\verb|emf|&=&-\frac{\Re\Big(i \varepsilon_{jmz} k_0 \widehat{\mathcal{E}_m}(k_0,t)\widehat{B_j}^*(k_0,t)\Big)}{|\widehat{B_j}(k_0,t)|},\\
\label{CorrelationE}
\verb|resistive term|&=&-\eta k_0^2|\widehat{B_j}(k_0,t)|,
\end{eqnarray}
$j$ being the magnetic field component studied ($j=x,y$) and $\varepsilon_{jmn}$ being the Levi-Civita tensor.
These quantities, with the phases of their associated terms in (\ref{bxbudgeteqn})-(\ref{bybudgeteqn}), are plotted in Fig.~\ref{bxbudget} for budget (\ref{bxbudgeteqn}) and in Fig.~\ref{bybudget} for budget (\ref{bybudgeteqn}).

We
find in Fig.~\ref{bxbudget} the long-timescale cycle we have already described. The shear term is positive in the beginning of the cycle, but becomes negative for $t>270$, whereas the EMF has a systematic resistive effect, with a clear phase correlation between the EMF and the resistive term. According to these results, the long-timescale cycle in the azimuthal field comes from the behaviour of the radial field, whereas the radial EMF may be seen, in first approximation, as a turbulent resistivity. As expected from this conclusion, $B_y$ is also oscillating on a long timescale (Fig.~\ref{bybudget}) with a smaller amplitude compared to $B_x$. Moreover, this cycle comes clearly from an oscillation of the azimuthal EMF $\mathcal{E}_x$, which is reversed at $t\sim243$. Note, however, that the EMF is very small ($\sim 5\times 10^{-4}$) compared to quantities in the budget for the azimuthal field, which may explain the long period of these oscillations. 

\begin{figure*}
   \centering
   \includegraphics[width=0.45\linewidth]{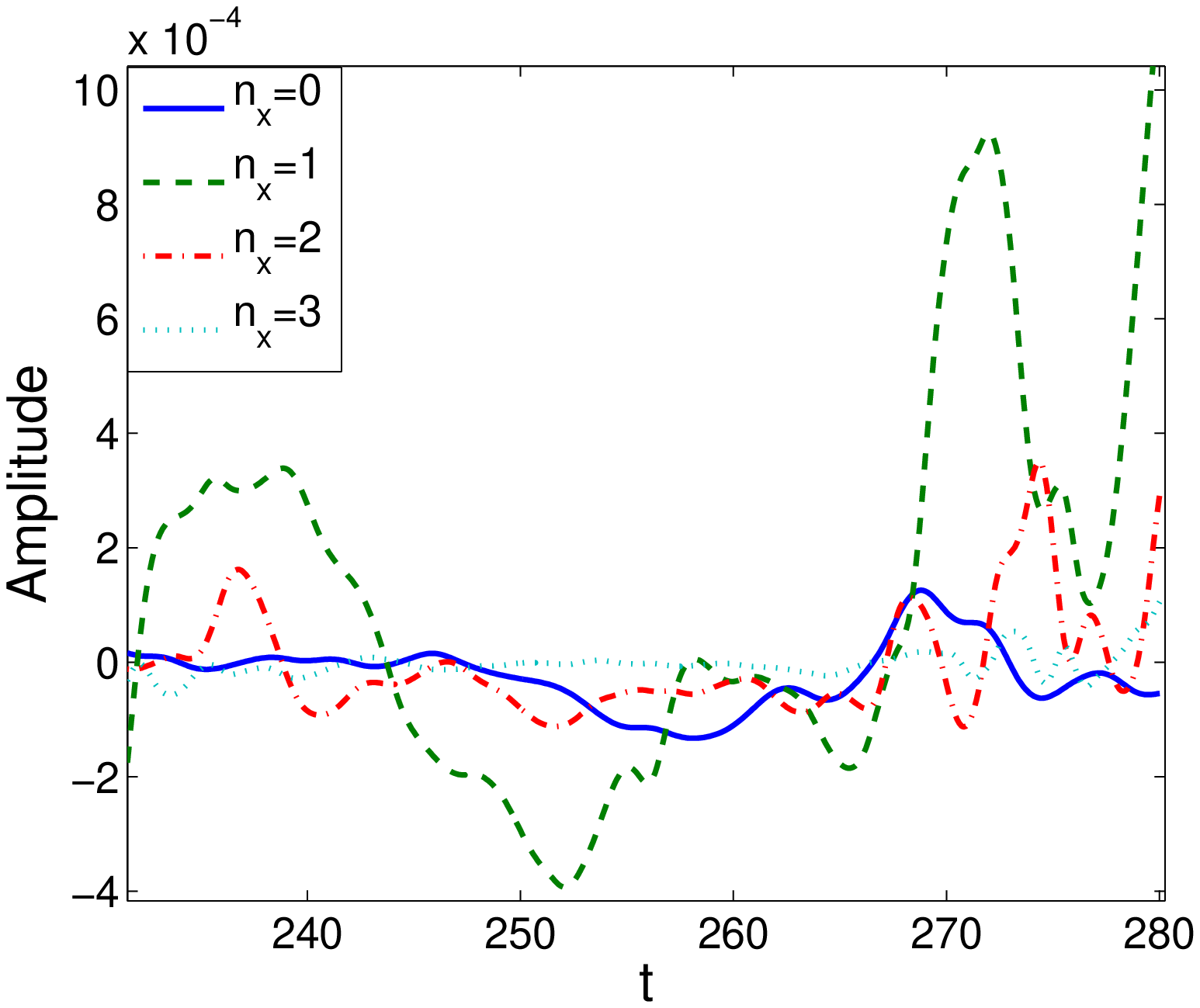}
   \includegraphics[width=0.45\linewidth]{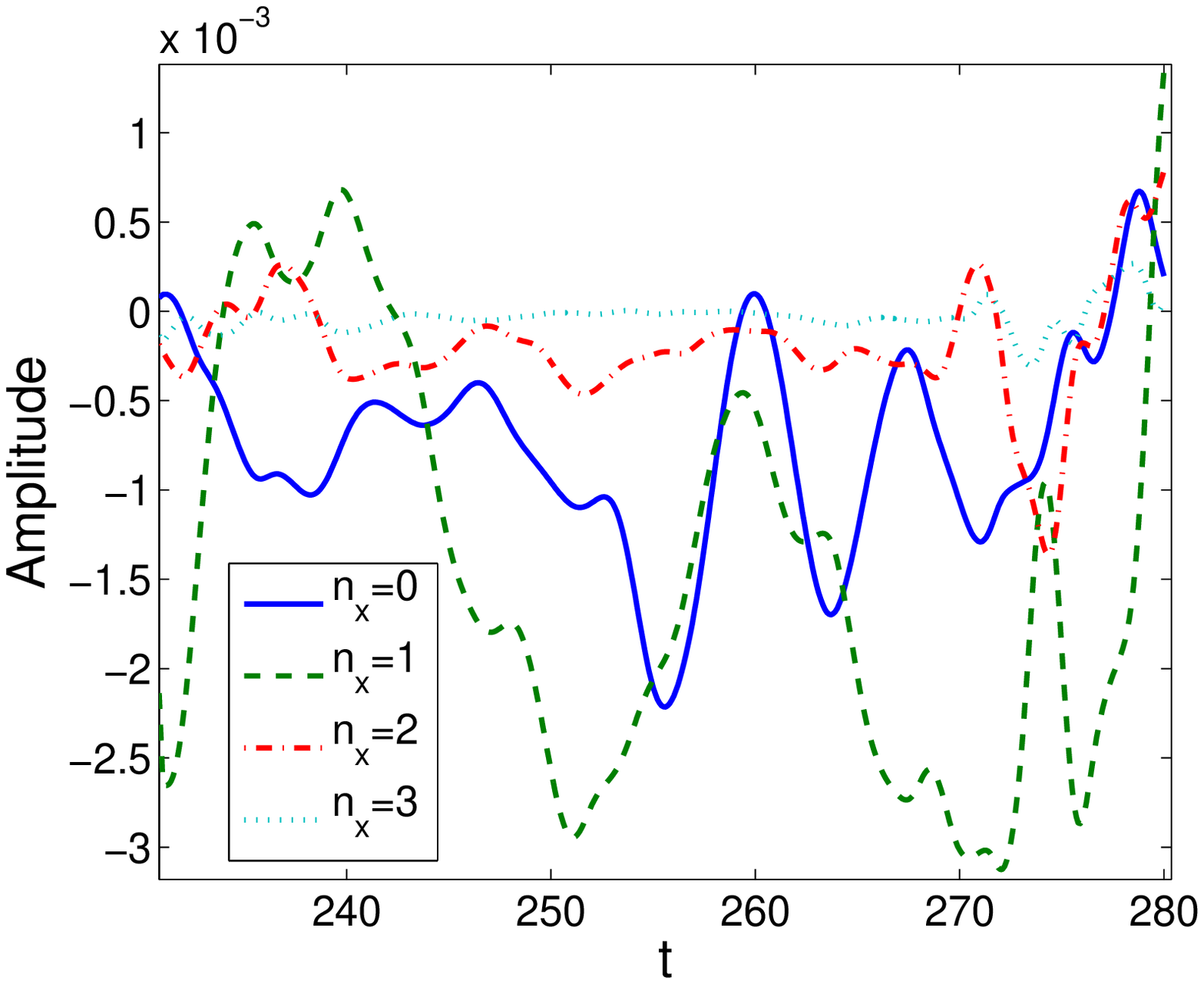}
   \caption{Contribution of nonaxisymmetric waves to the Fourier-transformed EMF $\widehat{\bm{\mathcal{E}}}(k_0)$. The largest nonaxisymmetric structures are responsible for $\mathcal{E}_x$ (left panel) whereas $\mathcal{E}_y$ involves both axisymmetric and nonaxisymmetric contributions (right panel).} 
              \label{enaxi}%
    \end{figure*}

\subsection{Modal analysis of non axisymmetric structures \label{EfieldSection}}
The EMFs described previously are obviously nonlinear terms, involving a coupling between $\bm{v}$ and $\bm{B}$.  An interesting question is therefore which modes contribute most to the EMFs observed in Figs \ref{bxbudget}--\ref{bybudget}. In particular, one may wonder whether nonaxisymmetric structures play a role and if so, which structures are dominant. Following this idea, we compute the following quantities:
\begin{eqnarray}
\nonumber \widehat{\bm{\mathcal{E}}}(n_x,k_0)=\frac{1}{L_y L_z}\int_0^{L_y} \! \! \! \!\int_0^{L_z} &&\!\!\!\!\!\! 2\Re[\widetilde{\bm{v}}(n_x,y,z)\bm{\times}\widetilde{\bm{B}}(n_x ,y,z)^*]\\&&\times \exp(-ik_0 z) \, dy \,dz,
\end{eqnarray}
with the fields $\widetilde{\varphi}(n_x,y,z)$ defined by the Fourier coefficient in the $x$ direction:
\begin{equation}
\widetilde{\varphi}(n_x,y,z)=\int_0^{L_x} \!\!\varphi(x,y,z)\exp\Big(-2i\pi n_x\frac{x}{L_x}\Big)\,dx.
\end{equation}
In this notation, $\widehat{\bm{\mathcal{E}}}(n_x,k_0)$ corresponds to the contribution of the nonaxisymmetric mode $n_x$ to the total EMF $\widehat{\bm{\mathcal{E}}}(k_0)$. We then put $\widehat{\bm{\mathcal{E}}}(n_x,k_0)$ in equation (\ref{emfdef}) and plot the contribution of the first few nonaxisymmetric modes in Fig.~\ref{enaxi}. From this figure, it is clear that most of the EMF comes from the largest azimuthal wavelengths, $n_x<3$. However, the detailed generation of $\mathcal{E}_x$ and $\mathcal{E}_y$ is somewhat different: although the contribution of axisymmetric modes ($n_x=0$) to $\mathcal{E}_x$ is small, these modes are clearly important for $\mathcal{E}_y$. Therefore, the azimuthal EMF $\mathcal{E}_x$ comes essentially from the largest nonaxisymmetric mode whereas the radial EMF comes from a more general effect of both nonaxisymmetric and axisymmetric structures. These results show that the magnetic cycles observed in the simulations are primarily due to three-dimensional structures, as one would expect from Cowling's antidynamo theorem \citep{C81}. 

\begin{figure}
   \centering
   \includegraphics[width=0.95\linewidth]{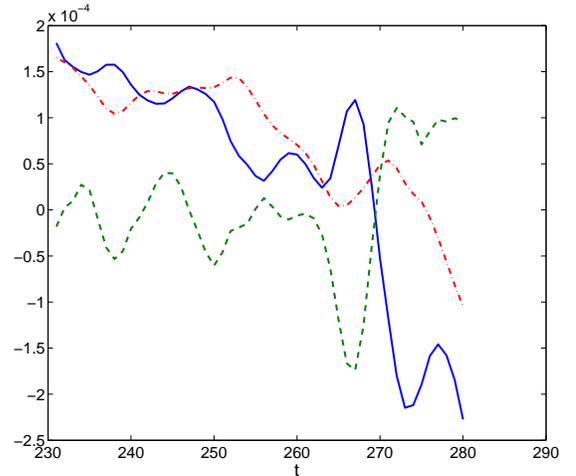}
      \caption{Magnetic helicity history during one cycle. The plain curve is the helicity due to the large-scale field $k_z=2\pi/L_z$, the dashed line corresponds to the helicity associated with the largest nonaxisymmetric modes, and the total helicity is plotted as a dash-dotted line.} 
              \label{helhistory}%
\end{figure}

It has been pointed out by \cite{B07} that monitoring the helicity evolution may be key to understanding large-scale field production. Moreover, since $\bm{\mathcal{E}\cdot B}$ is non-zero for the largest vertical modes, some large-scale magnetic helicity should be produced. To explore the helicity behaviour in our case,  we compare the box-averaged magnetic helicity due to the vertical mode $\bm{k}=(2\pi/L_z)\bm{e_z}$ with the magnetic helicity associated with the largest nonaxisymmetric modes $n_\phi=1,2$ in Fig.~\ref{helhistory}. During one cycle, we note that the helicity associated with the vertical modes changes sign when $B_y$ is reversed ($t\simeq 270$). Moreover, this reversal seems to be associated with an exchange of magnetic helicity between the nonaxisymmetric and the axisymmetric vertical modes. Note however that the total magnetic helicity also varies significantly during one cycle. Since the box averaged total magnetic helicity is conserved in ideal MHD, any variation of the total helicity must be due to a resistive (and therefore small-scale) process. In that way, we cannot make a very clear distinction between the helicity exchanged with the nonaxisymmetric modes, and the helicity due to small scale production without any further analysis.  Moreover, we emphasise that the averaged helicity appears to be about $10^{-2}$ times smaller than $L_z\langle b^2\rangle$, which is probably a consequence of the numerous symmetries in unstratified shearing boxes. Therefore, the helicity produced in our simulations might be more a consequence of the underlying nonlinear dynamo process rather than its cause.

In this section, we have shown that zero-net-flux MRI simulations exhibit long-timescale oscillations in the azimuthal magnetic field $B_x$ with a large vertical wavelength. Studying the budget of $B_x$, it appears that these cycles are primarily an amplification of oscillations in $B_y$ through the shear term, whereas the EMF $\mathcal{E}_y$ acts as a turbulent resistivity. A similar analysis of the $B_y$ cycles shows that they are generated by a long-period but small-amplitude oscillation of $\mathcal{E}_x$. Finally, we used a modal analysis to demonstrate that $\mathcal{E}_x$ and $\mathcal{E}_y$ are mostly generated by large-scale nonaxisymmetric waves, making these cycles a fully three-dimensional problem. 

Therefore, to have a better understanding of this phenomenological picture, one needs to study these nonaxisymmetric structures and the EMFs associated with them, which is the topic of the next section.

\section{Linear analysis of nonaxisymmetric waves}
\subsection{Model and equations}
To understand the nonaxisymmetric origin of the EMF described previously, we consider a linear model including shearing waves in the presence of a background azimuthal magnetic field with a vertical structure, $B_x^0(z)$. This vertical structure is required in order to compare the EMF generated by the perturbation with the background magnetic field. In this sense, this calculation differs from the initial MRI studies with a uniform azimuthal field \citep{BH92}.

Because of the mean shear, we cannot use a classical Fourier decomposition for nonaxisymmetric waves. Therefore, following \cite{GL65} and \cite{BH92}, we define a sheared frame comoving with the laminar flow:
\begin{equation}
x'=x-Syt\quad \quad y'=y \quad\quad z'=z\quad\quad t'=t.
\end{equation}
In this Lagrangian frame, we look for 
plane-wave solutions  $\phi=\phi(z',t') \exp[i(k'_xx'+k'_yy')]$ which may be transformed into the unsheared frame as:
\begin{equation}
\phi(x,y,z,t)=\bar{\phi}(z,t)\exp[i(k_xx+k_y(t)y)],
\end{equation}
with $k_x=k_x'$ and $k_y(t)=k_y'-Sk_xt$. We then consider perturbations from the background magnetic structure in the form of these shearing waves. This vertical structure is supposed to mimic the large-scale field observed in numerical simulations, i.e.\ being a function of $z$ only. Moreover, since the cycles have a long timescale compared to the shear, we also assume to a first approximation that this large-scale field is constant for a shearing wave. We therefore write the general linearized solution as:
\begin{eqnarray}
\bm{v}&=&\bm{\bar{v}}(z,t)\exp[i(k_xx+k_y(t)y)],\\
\bm{B}&=&\bm{\bar{b}}(z,t)\exp[i(k_xx+k_y(t)y)]+B_x^0(z)\bm{e_x},
\end{eqnarray}
where $|\bm{\bar{v}}|$ and  $|\bm{\bar{b}}|$ are supposed infinitely small.
Using these solutions in the evolution equations (\ref{motion})--({\ref{induction}), one eventually finds:
\begin{eqnarray}
\nonumber
\partial_t \bm{\bar{v}}&=&-(i\bm{k}+\bm{e_z}\partial_z)\bar{\Pi}+(2\Omega-S)\bar{v}_y\bm{e_x}-2\Omega \bar{v}_x \bm{e_y}\\
\label{motionlin}& &+ik_xB_x^0(z)\bm{\bar{b}} +\bar{b}_z\partial_zB_x^0(z)\bm{e_x}+\nu(\partial_z^2-k^2)\bm{\bar{v}},\\
\label{inductionlin}\partial_t \bm{\bar{b}}&=&S\bar{b}_y\bm{e_x}+ik_xB_x^0(z)\bm{\bar{v}}-\bar{v}_z\partial_zB_x^0(z)\bm{e_x}+\eta(\partial_z^2-k^2)\bm{\bar{b}},
\end{eqnarray}
where the generalised pressure is constrained by the incompressibility condition (\ref{Vstruct}) and $k^2=k_x^2+k_y^2(t)$. 

\begin{figure*}
   \centering
   \includegraphics[width=0.24\linewidth]{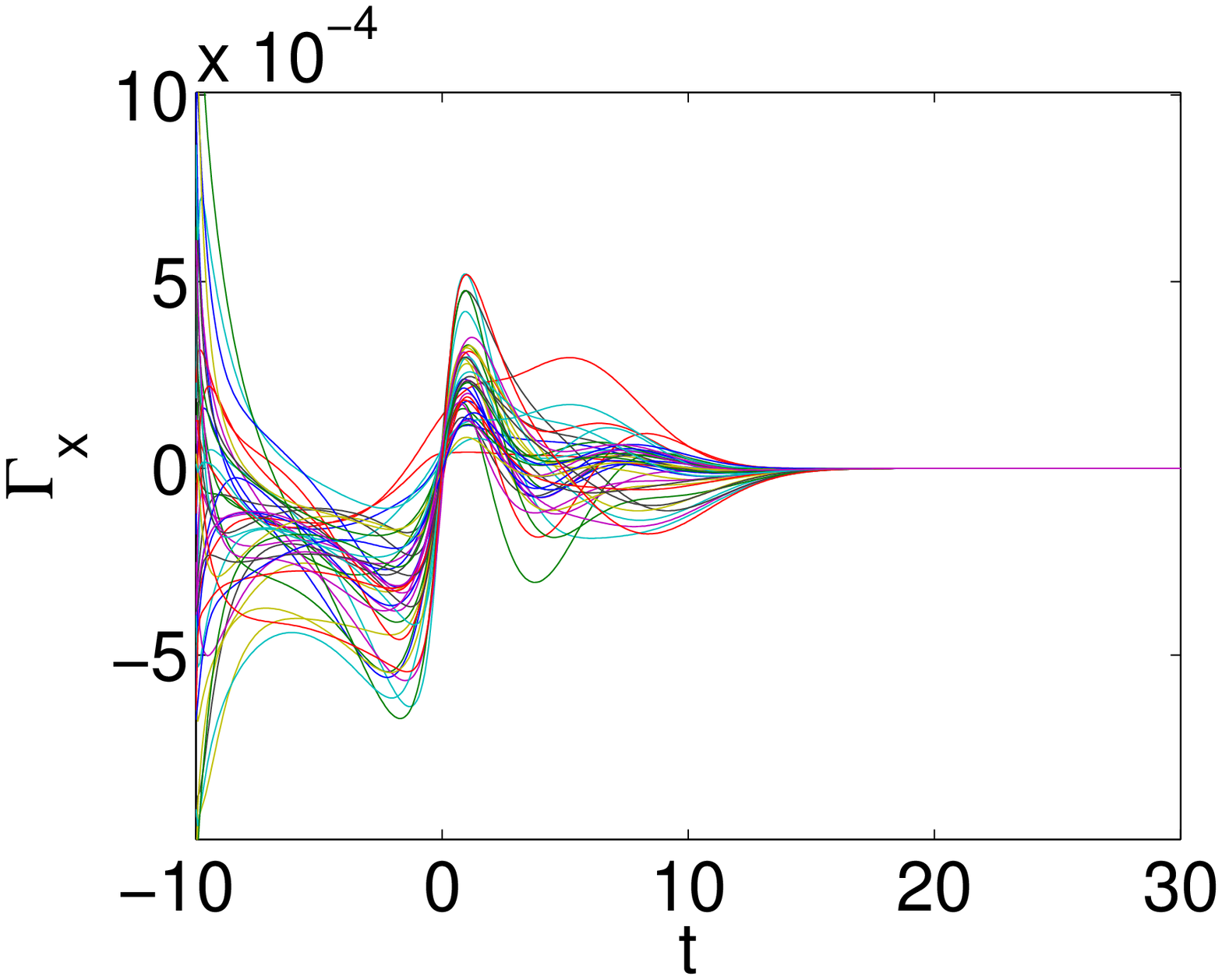}
   \includegraphics[width=0.24\linewidth]{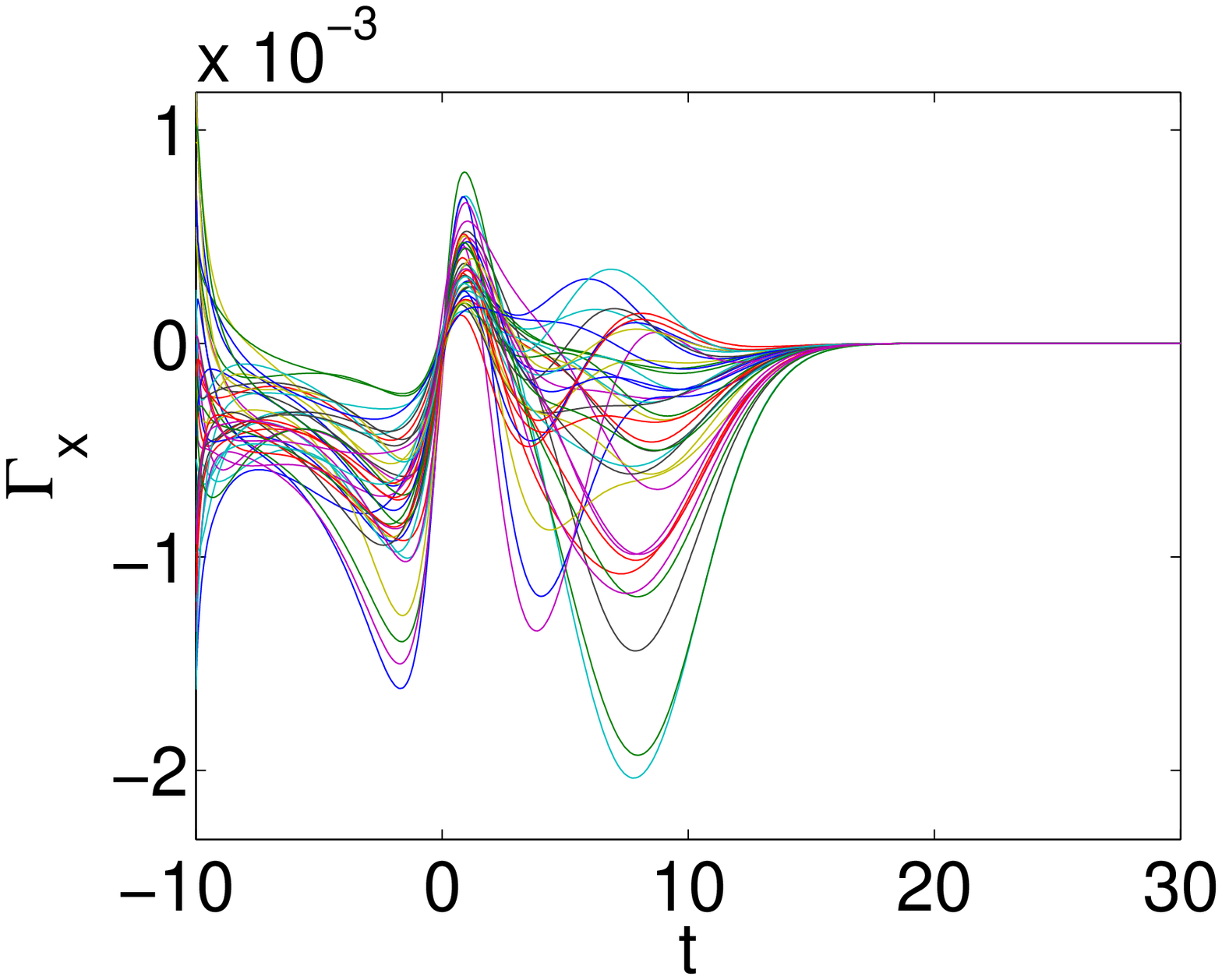}
   \includegraphics[width=0.24\linewidth]{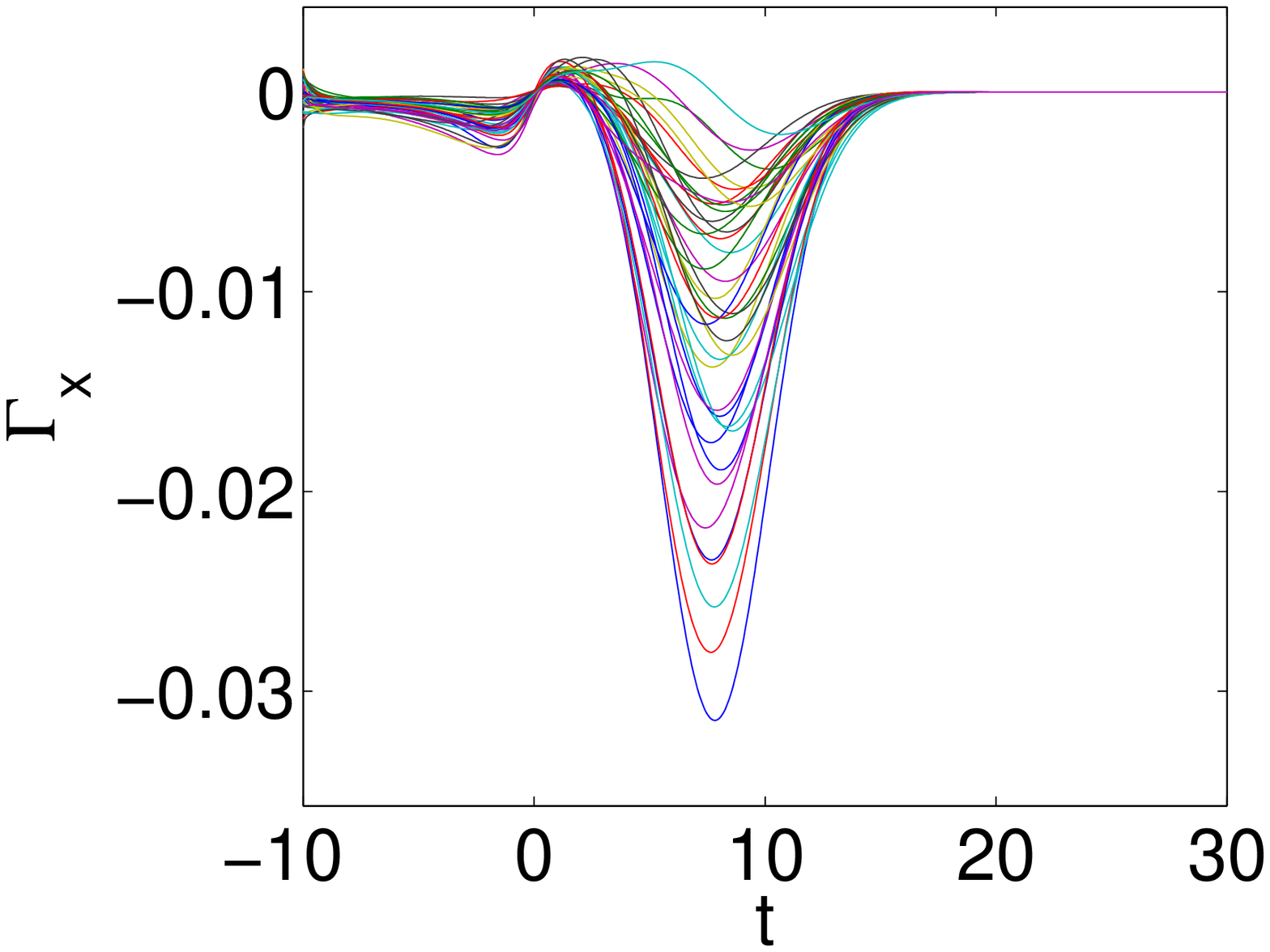}
   \includegraphics[width=0.24\linewidth]{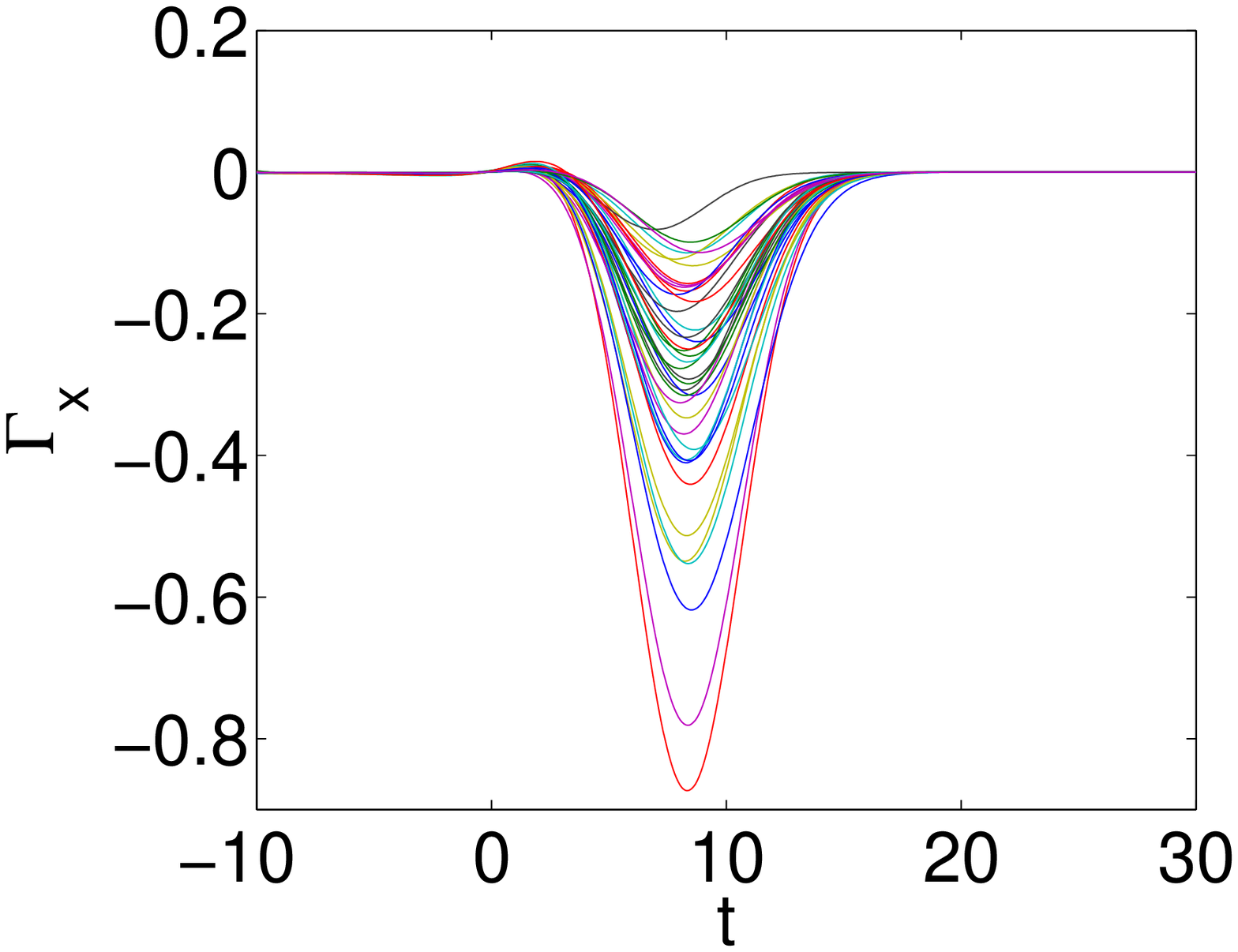}\\
   \includegraphics[width=0.24\linewidth]{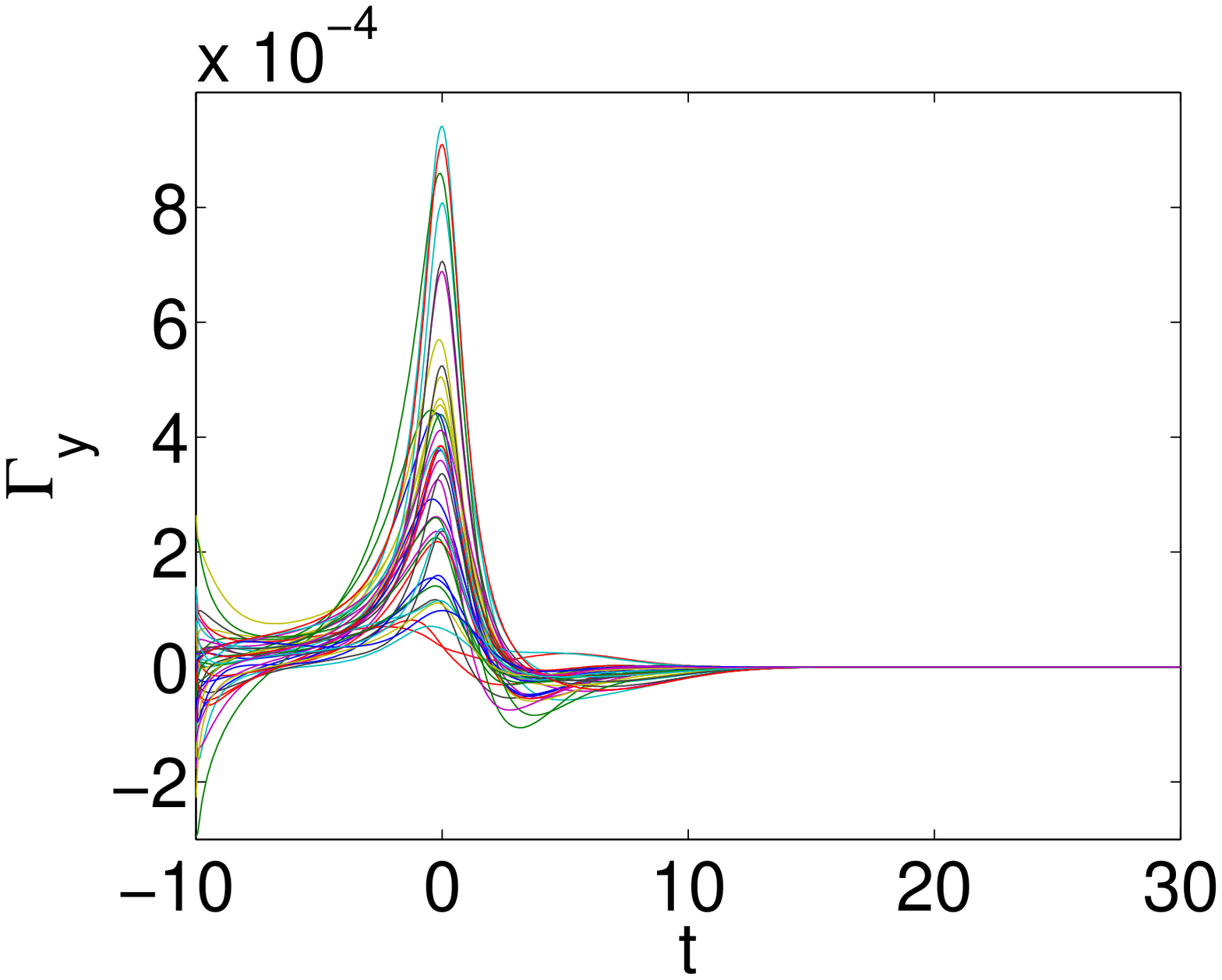}
   \includegraphics[width=0.24\linewidth]{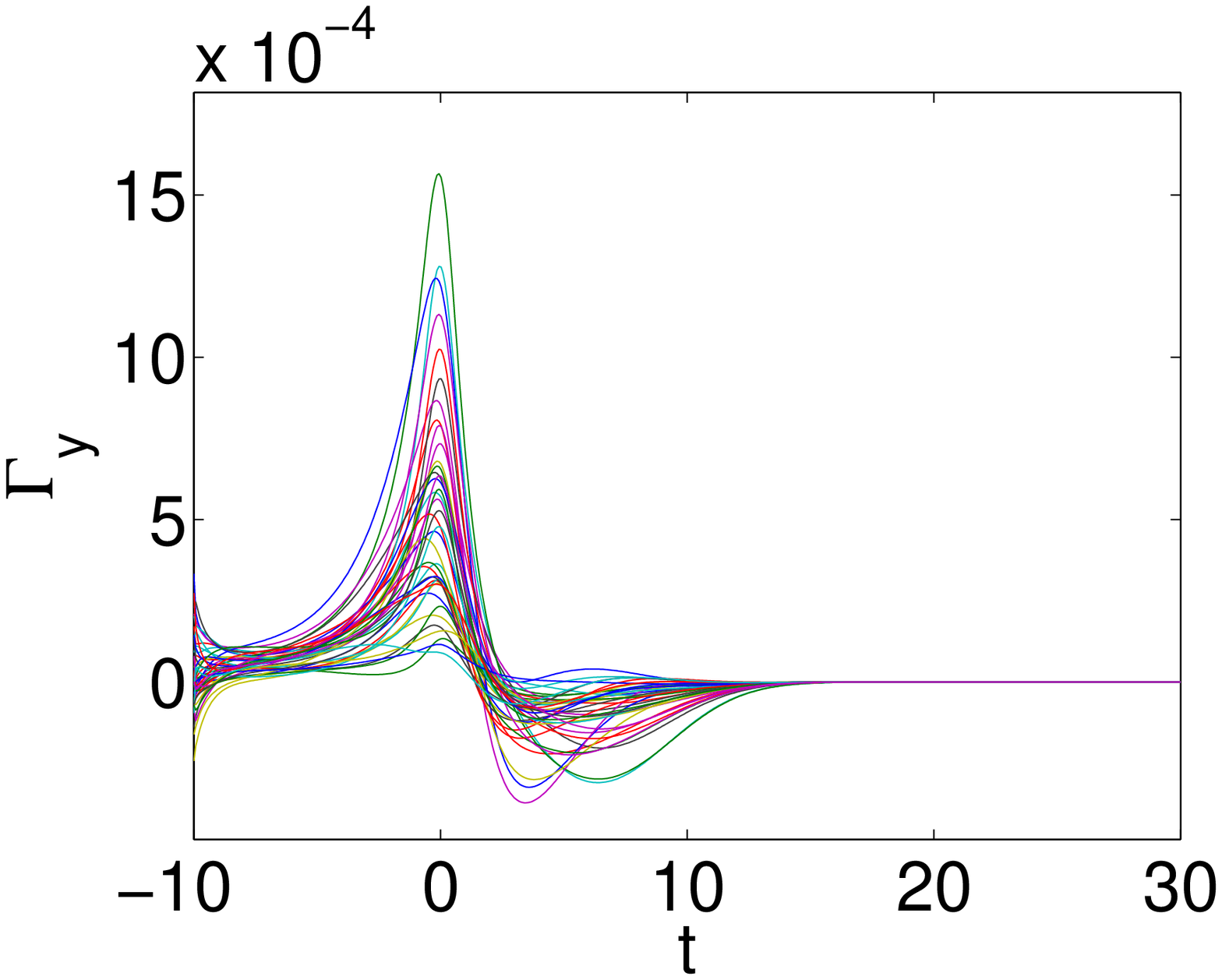}
   \includegraphics[width=0.24\linewidth]{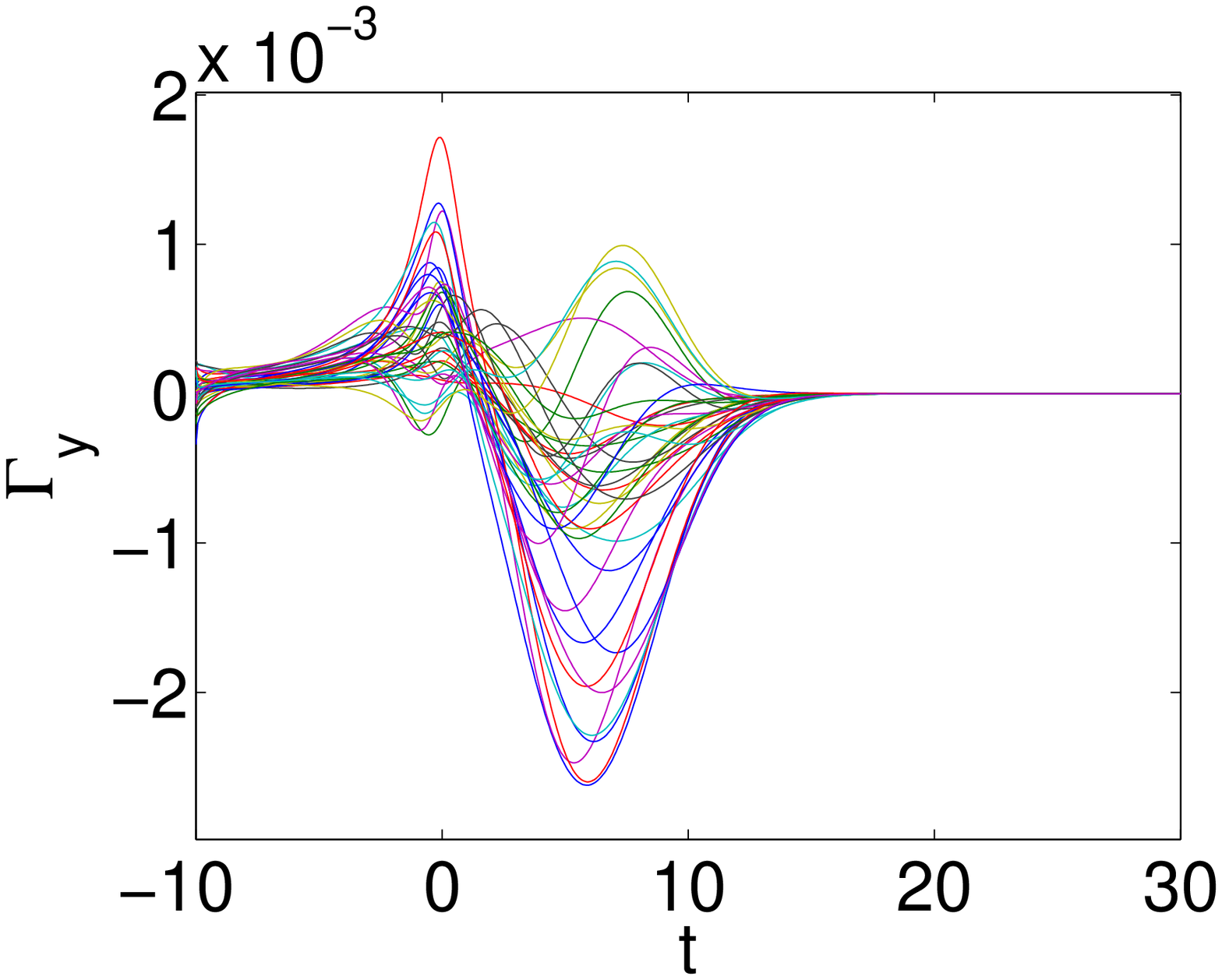}
   \includegraphics[width=0.24\linewidth]{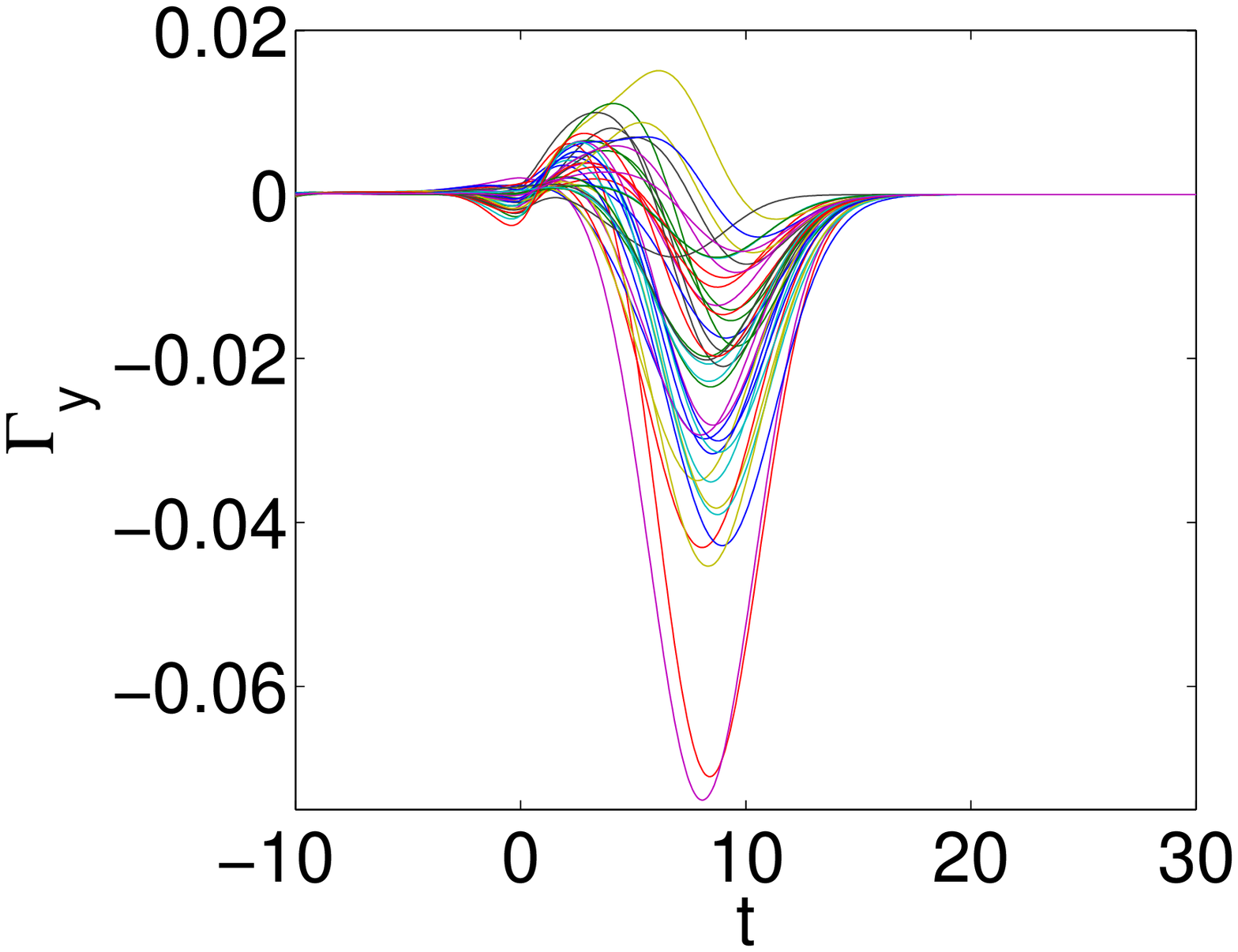}\\
   
   \caption{Evolution of $\Gamma_x$ (upper panels) and $\Gamma_y$ (lower panels) for a set of 40 shearing waves with random initial conditions, $Re=1600$ and $Rm=6400$. From left to right, the large-scale field is increased with $B_0=0.02;0.04;0.08;0.2$. $\langle \Gamma_x\rangle$ is always negative whereas $\langle\Gamma_y\rangle$ is positive for $B_0\lesssim 0.08$ and is reversed for larger field strengths. This is consistent with the cycles observed in the previous section (see text).} 
              \label{GammaPlotLRe}%
    \end{figure*}
    
\begin{figure*}
   \centering
   \includegraphics[width=0.24\linewidth]{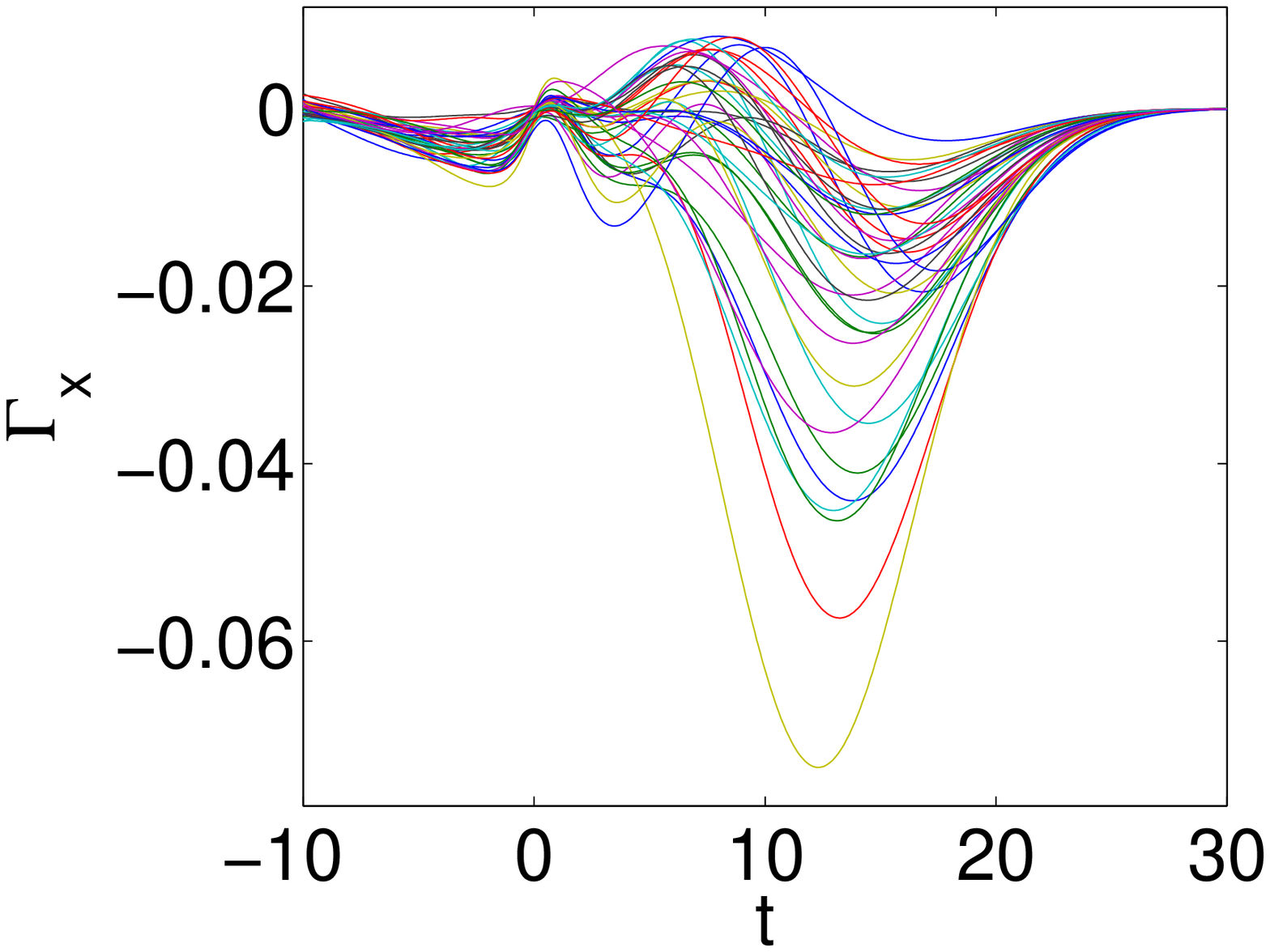}
   \includegraphics[width=0.24\linewidth]{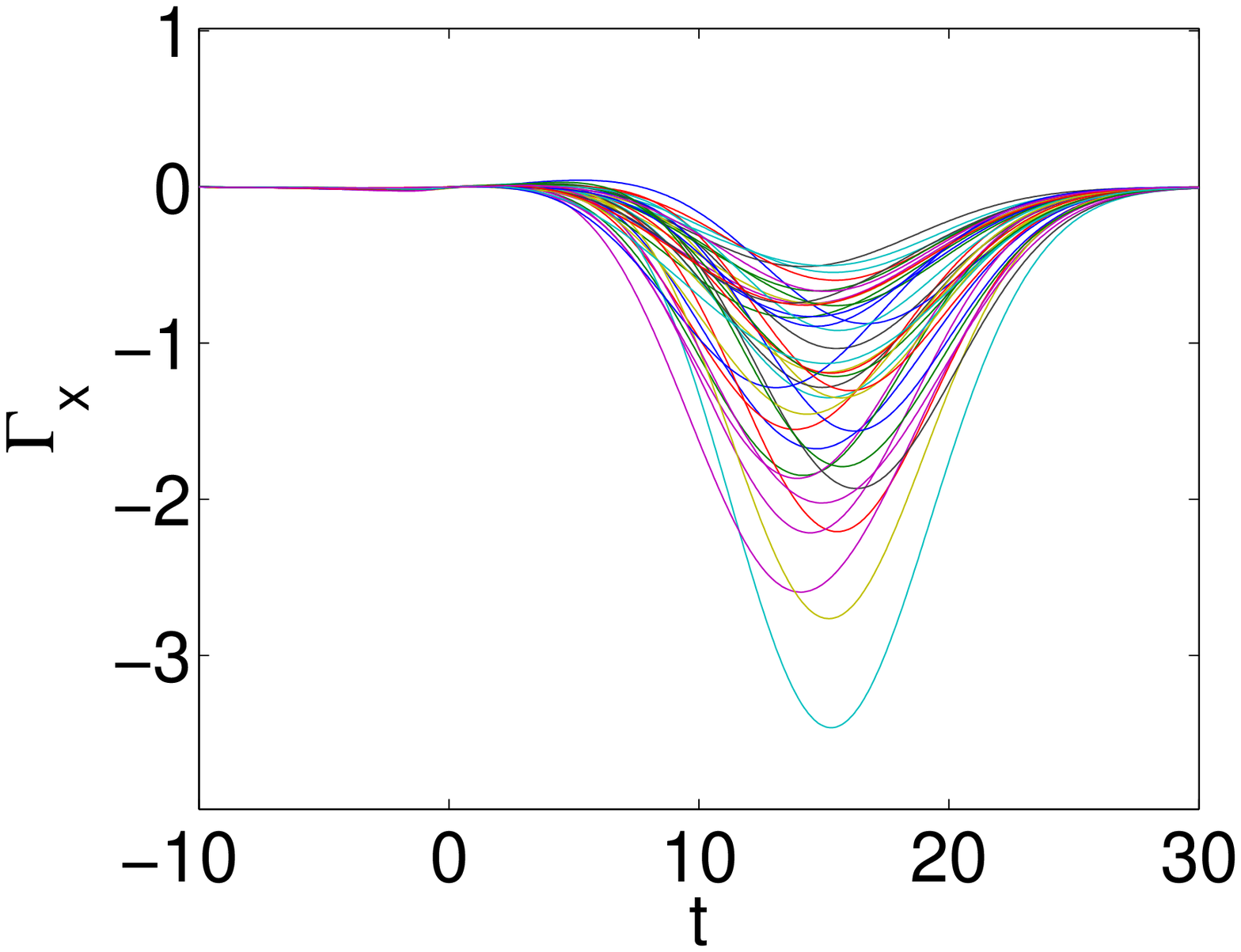}
   \includegraphics[width=0.24\linewidth]{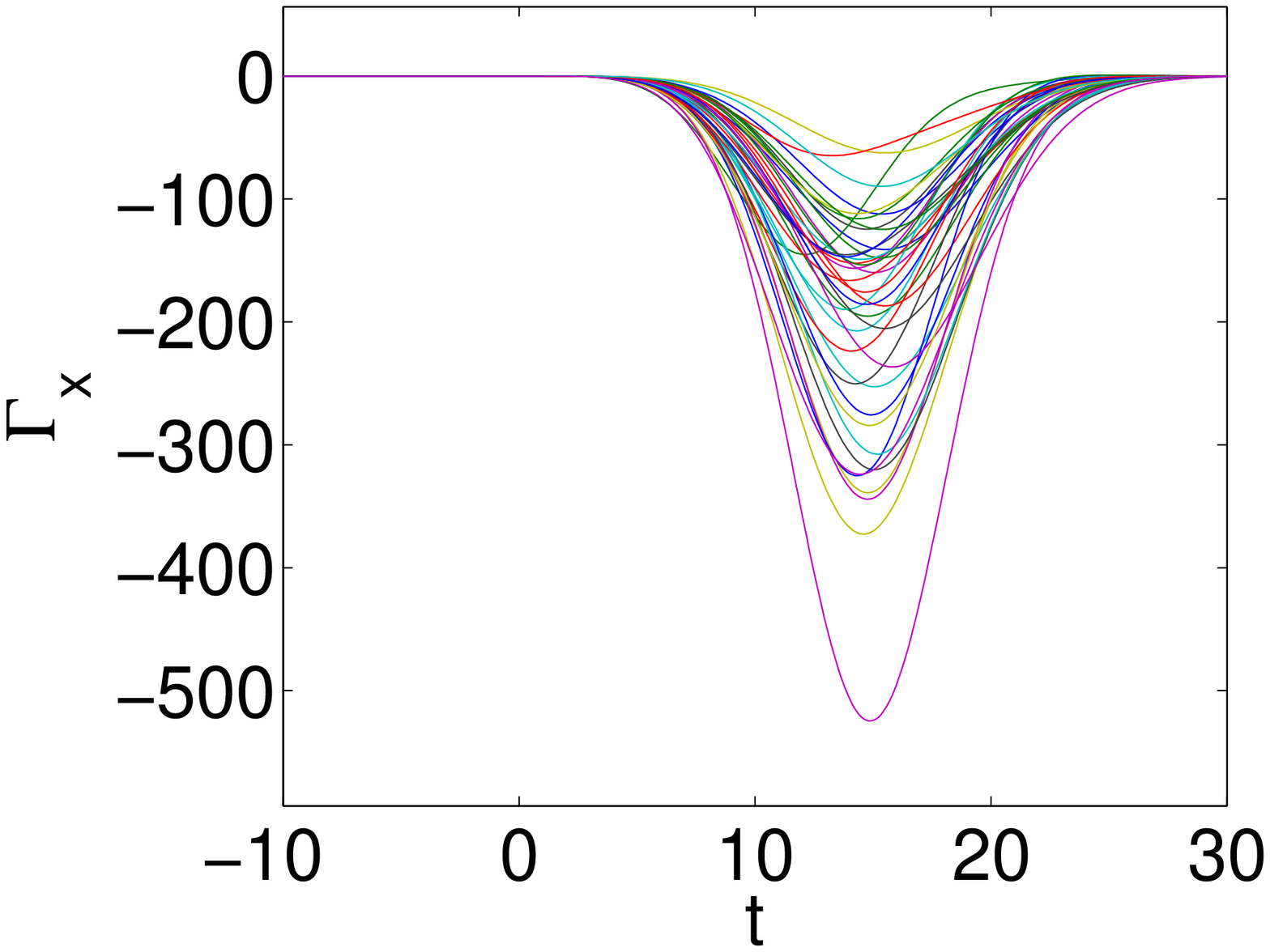}
   \includegraphics[width=0.24\linewidth]{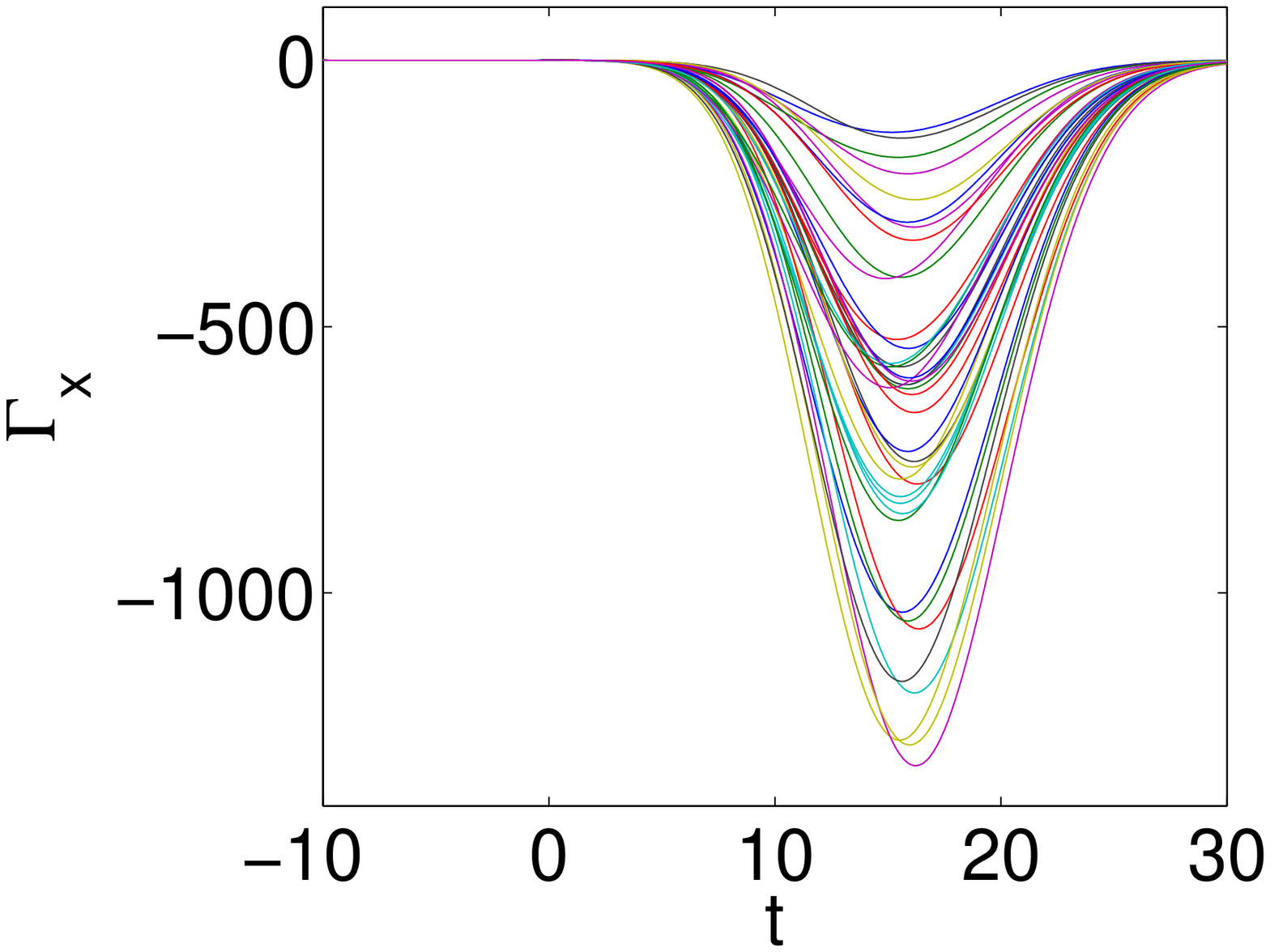}\\
   \includegraphics[width=0.24\linewidth]{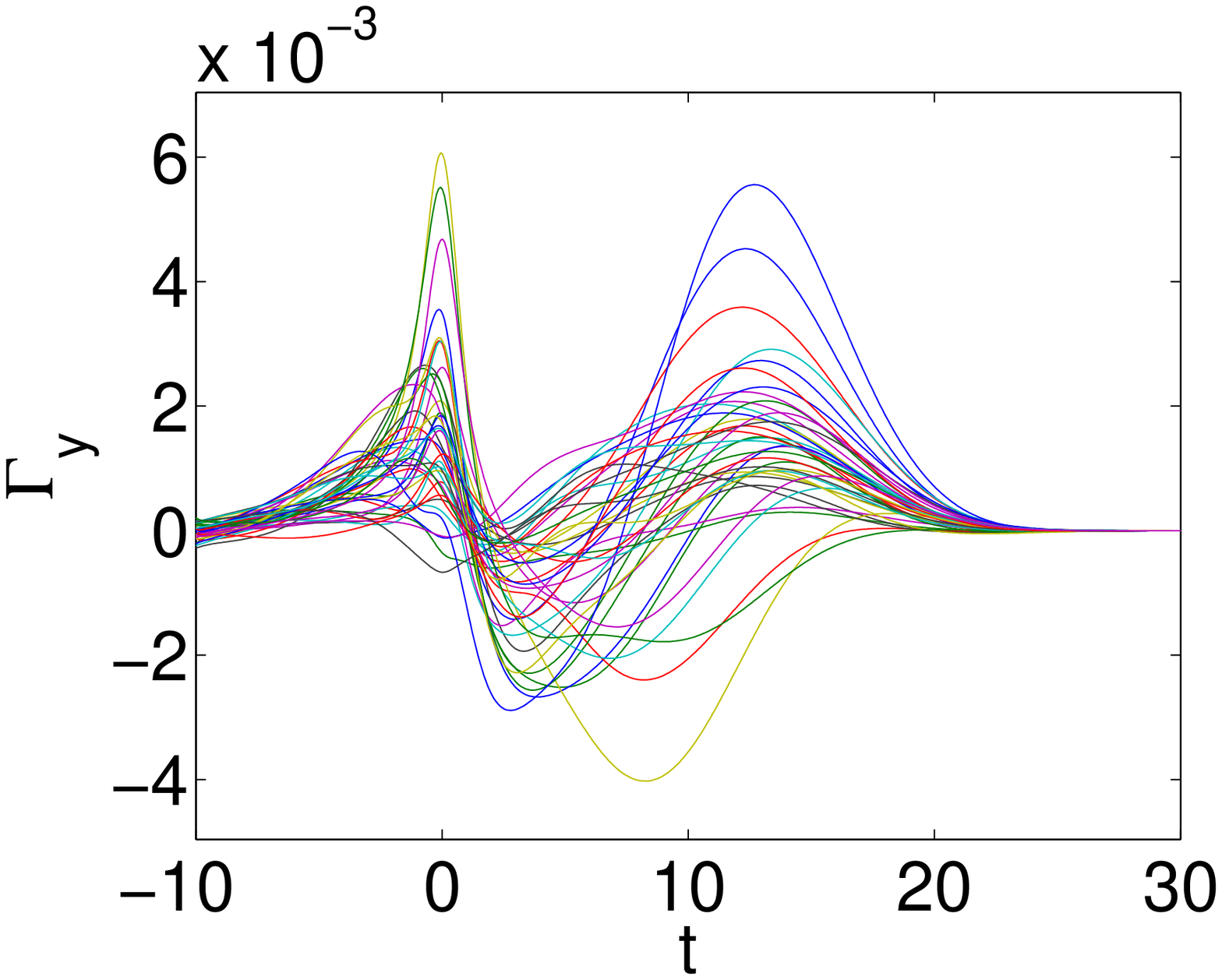}
   \includegraphics[width=0.24\linewidth]{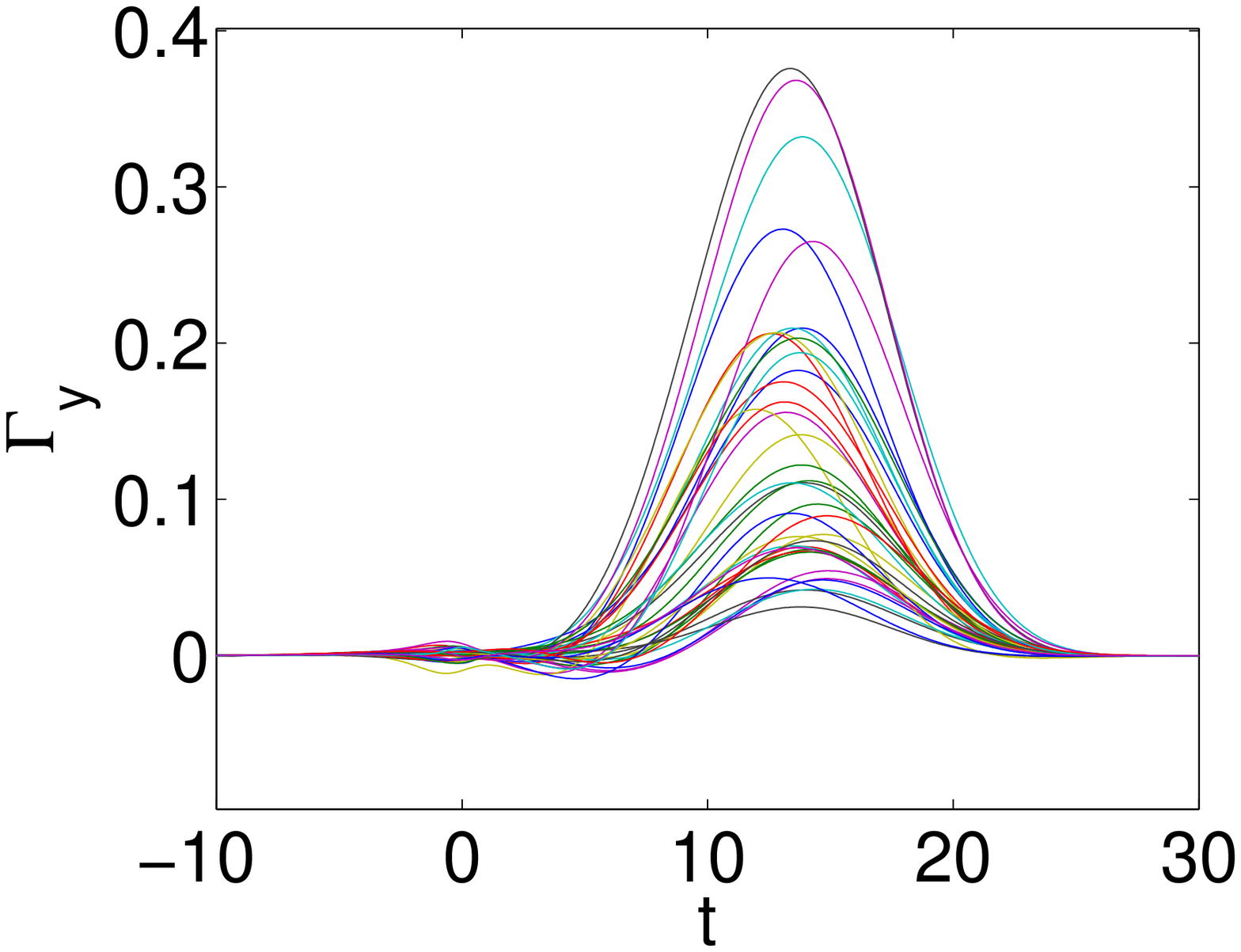}
   \includegraphics[width=0.24\linewidth]{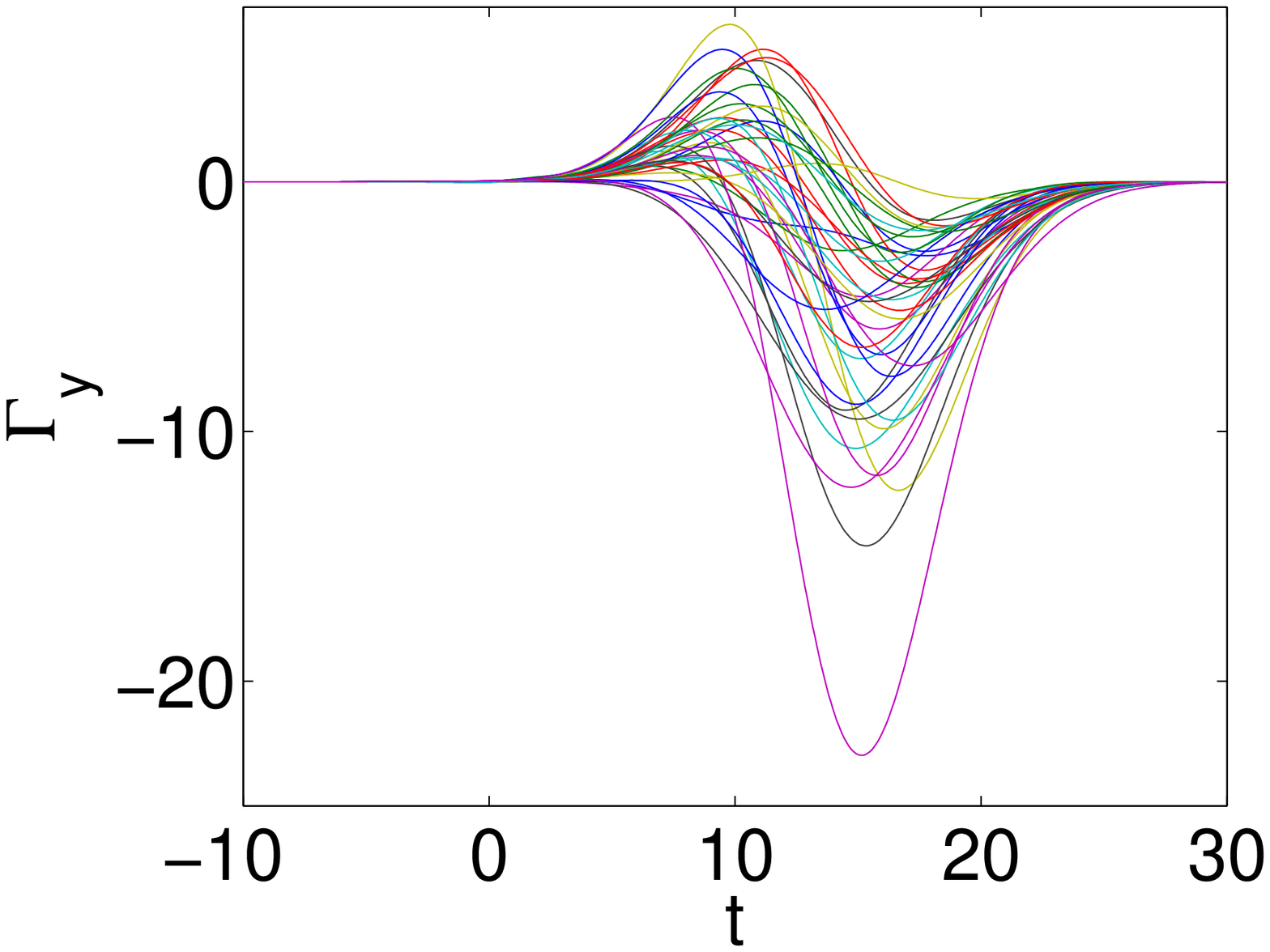}
   \includegraphics[width=0.24\linewidth]{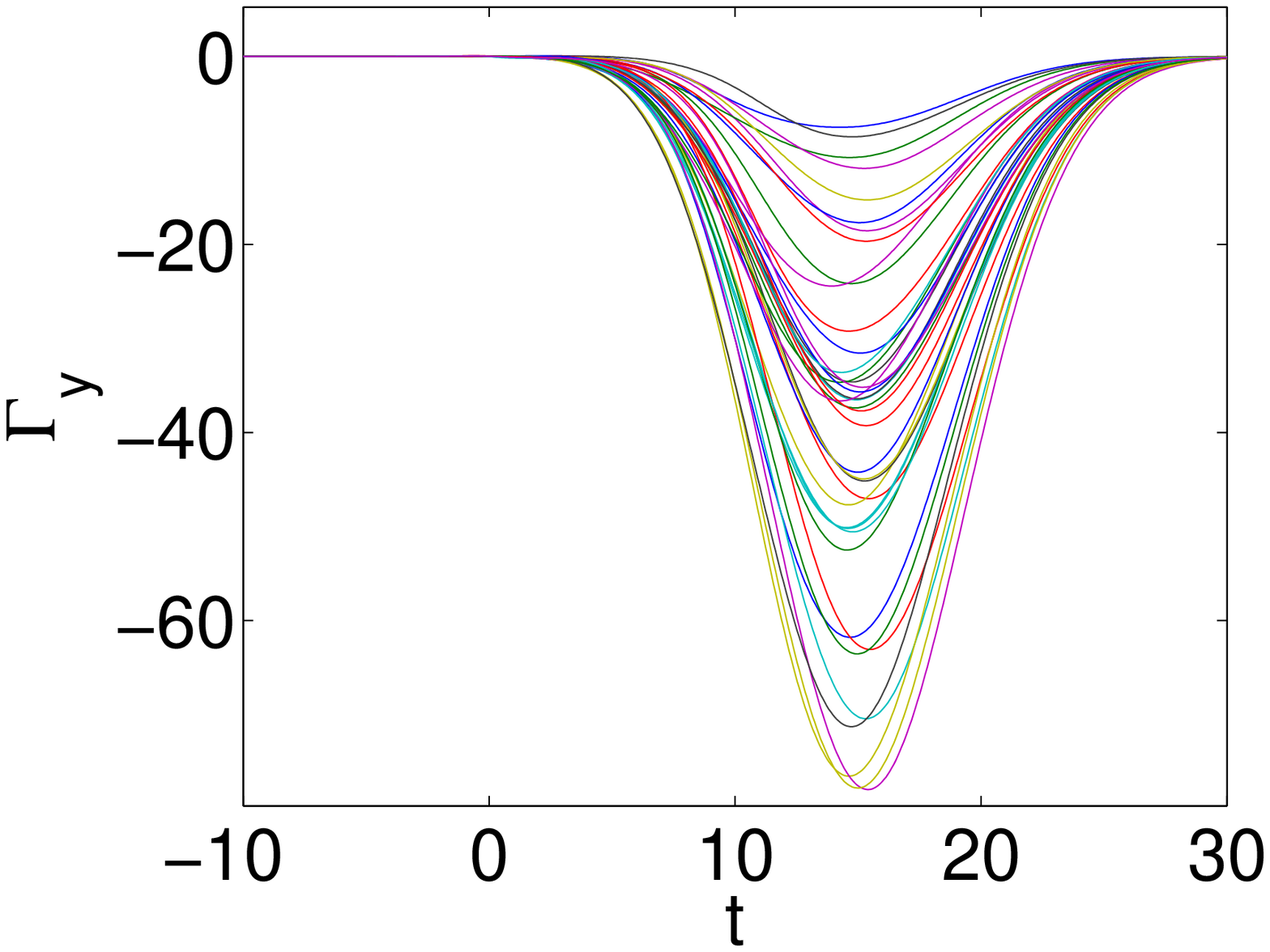}\\
   
   \caption{Evolution of $\Gamma_x$ (upper panels) and $\Gamma_y$ (lower panels) for a set of 40 shearing waves with random initial conditions, $Re=6400$ and $Rm=25600$. From left to right, the large-scale field is increased with $B_0=0.04;0.08;0.2;0.3$. As for Fig.~\ref{GammaPlotLRe}, $\langle \Gamma_x\rangle$ is always negative whereas  $\langle\Gamma_y\rangle$ is reversed in this case for $B_0\gtrsim 0.2$.} 
              \label{GammaPlotHRe}%
    \end{figure*}

\begin{figure*}
   \centering
   \includegraphics[width=0.24\linewidth]{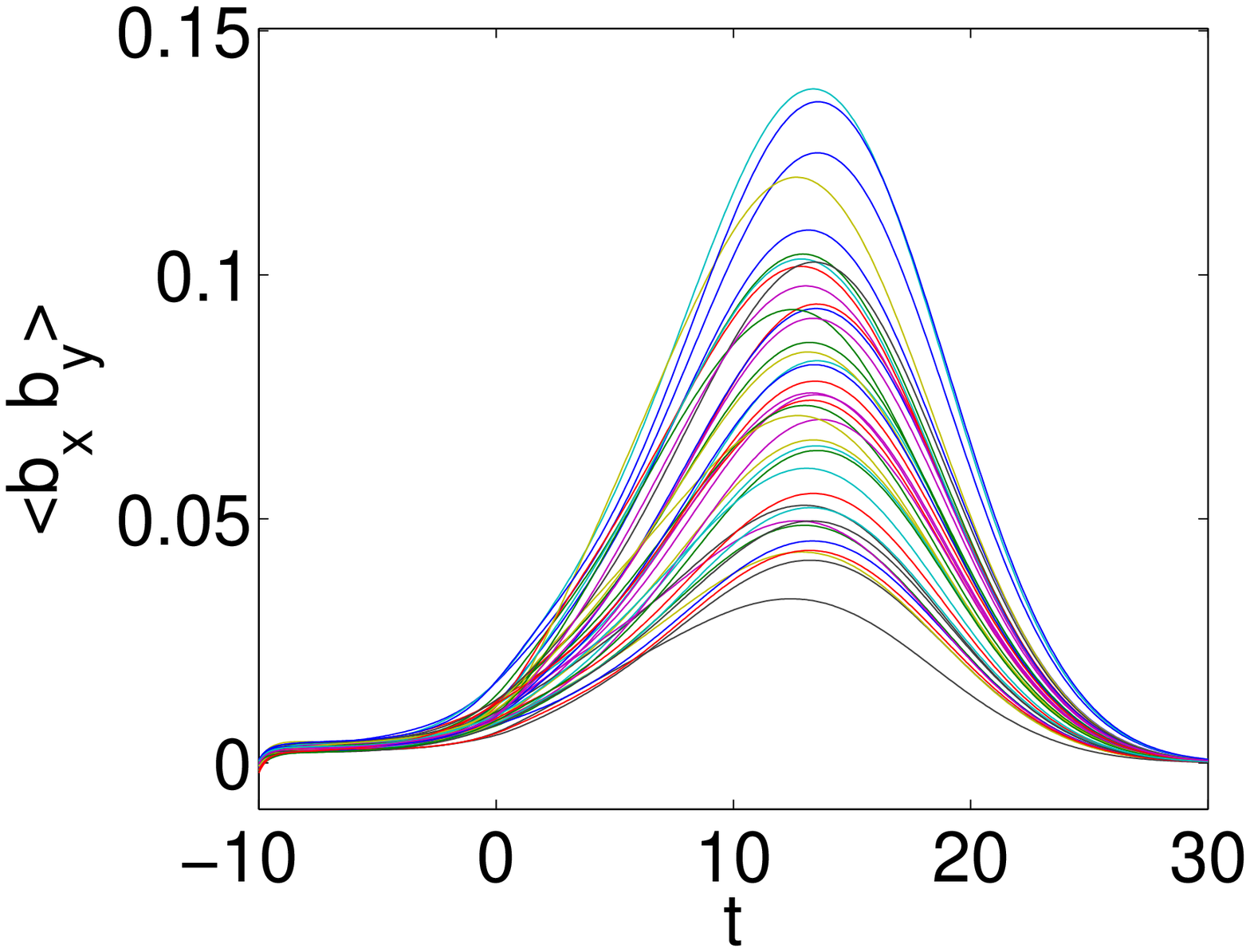}
   \includegraphics[width=0.24\linewidth]{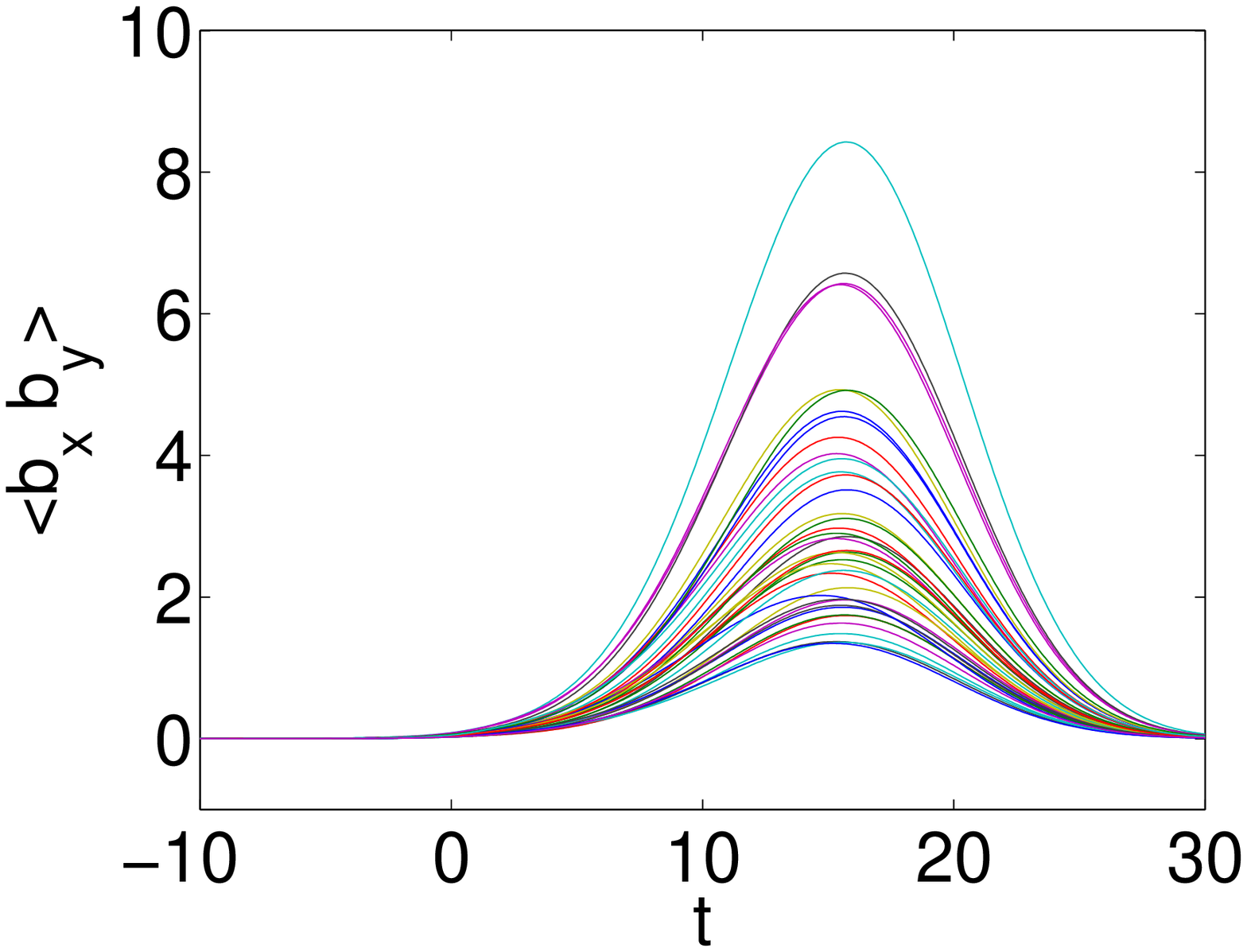}
   \includegraphics[width=0.24\linewidth]{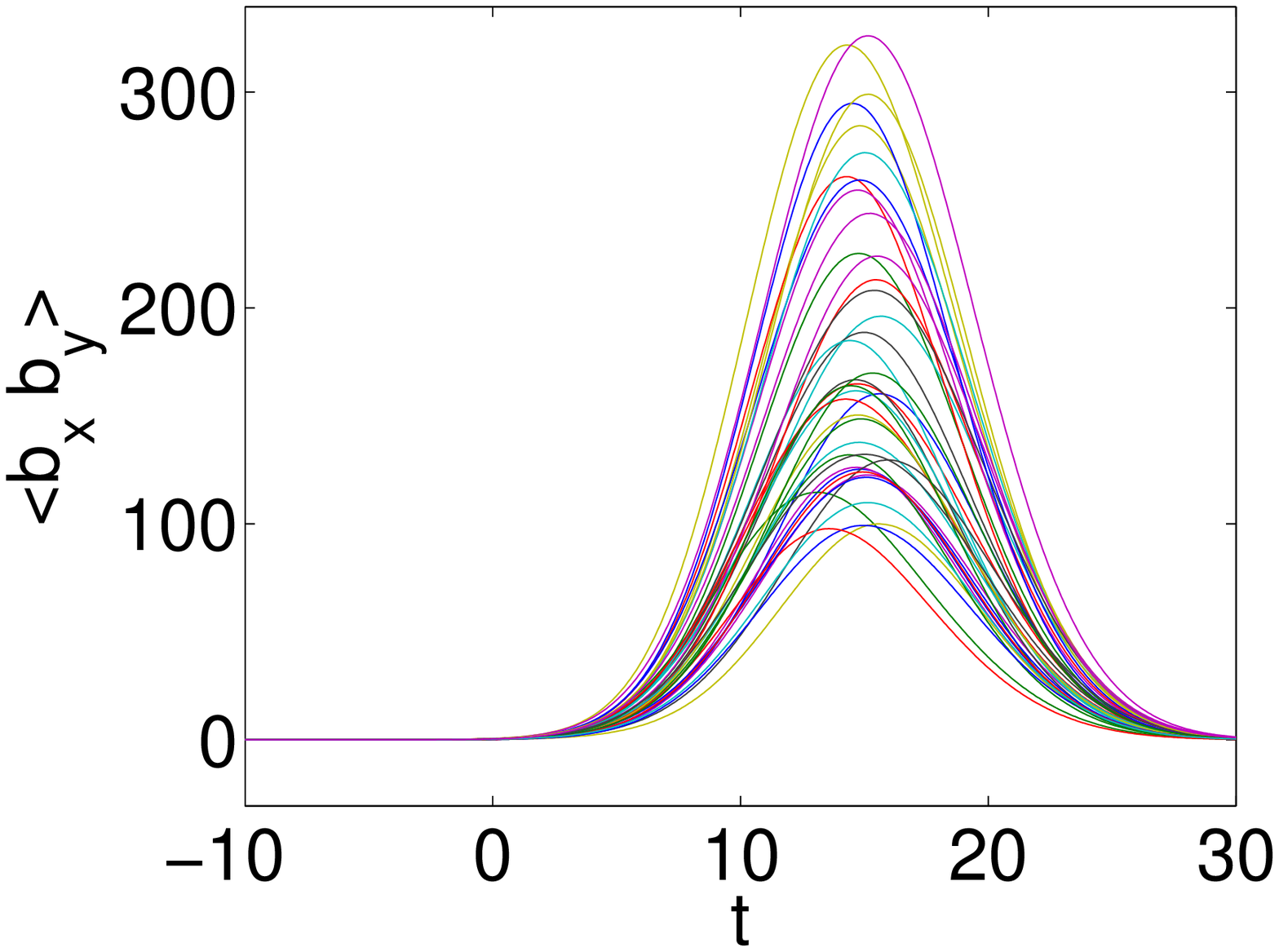}
   \includegraphics[width=0.24\linewidth]{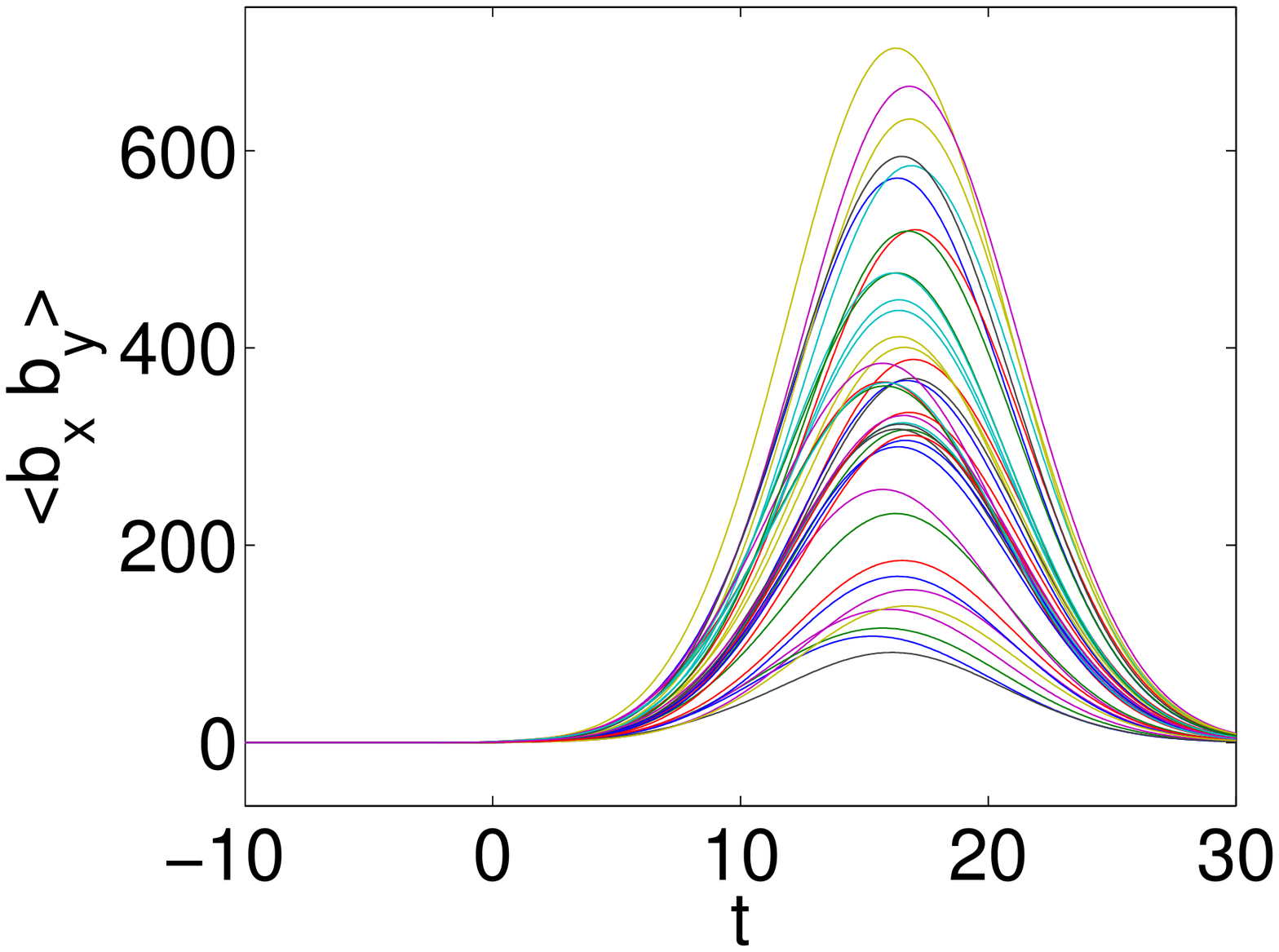}\\
   \includegraphics[width=0.24\linewidth]{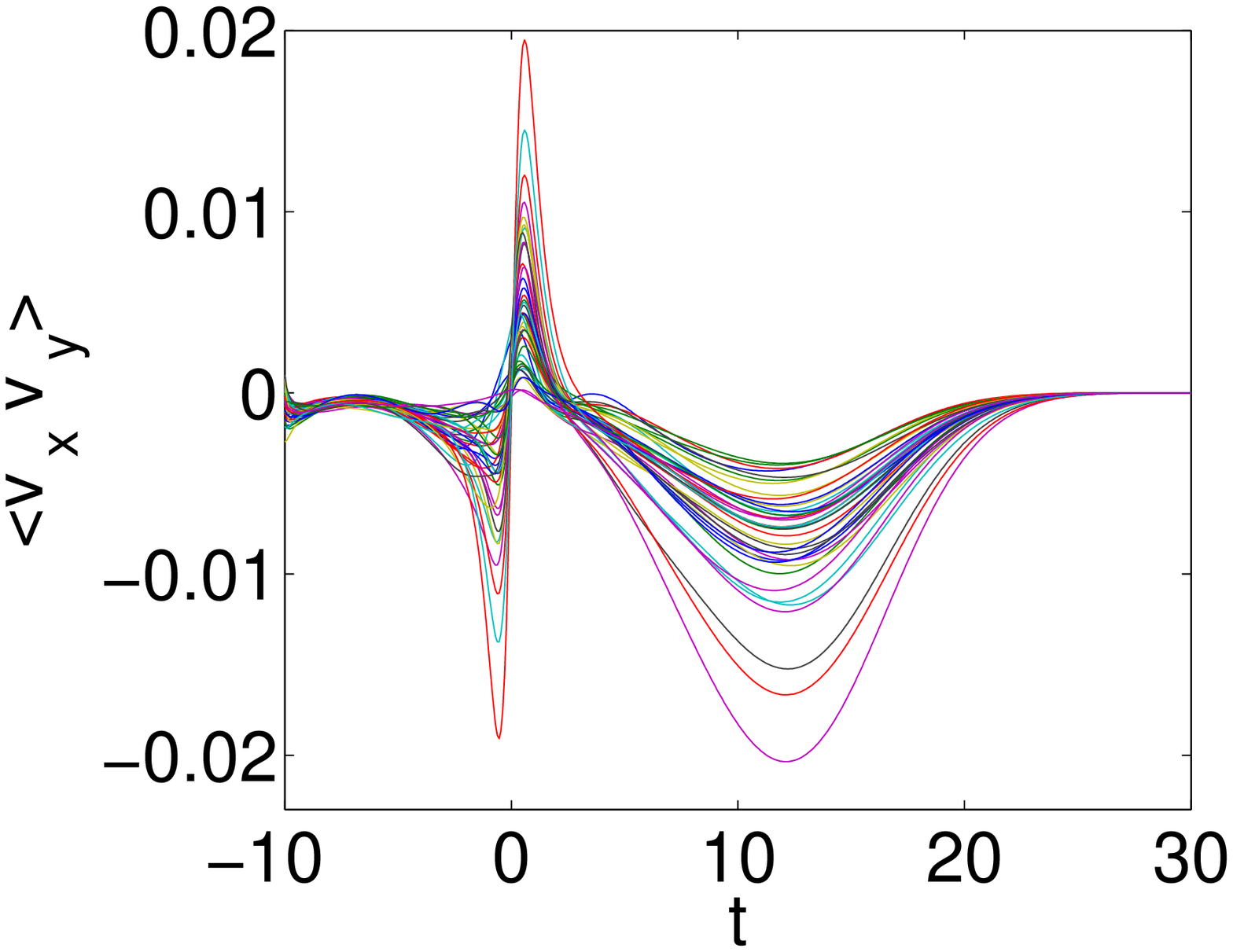}
   \includegraphics[width=0.24\linewidth]{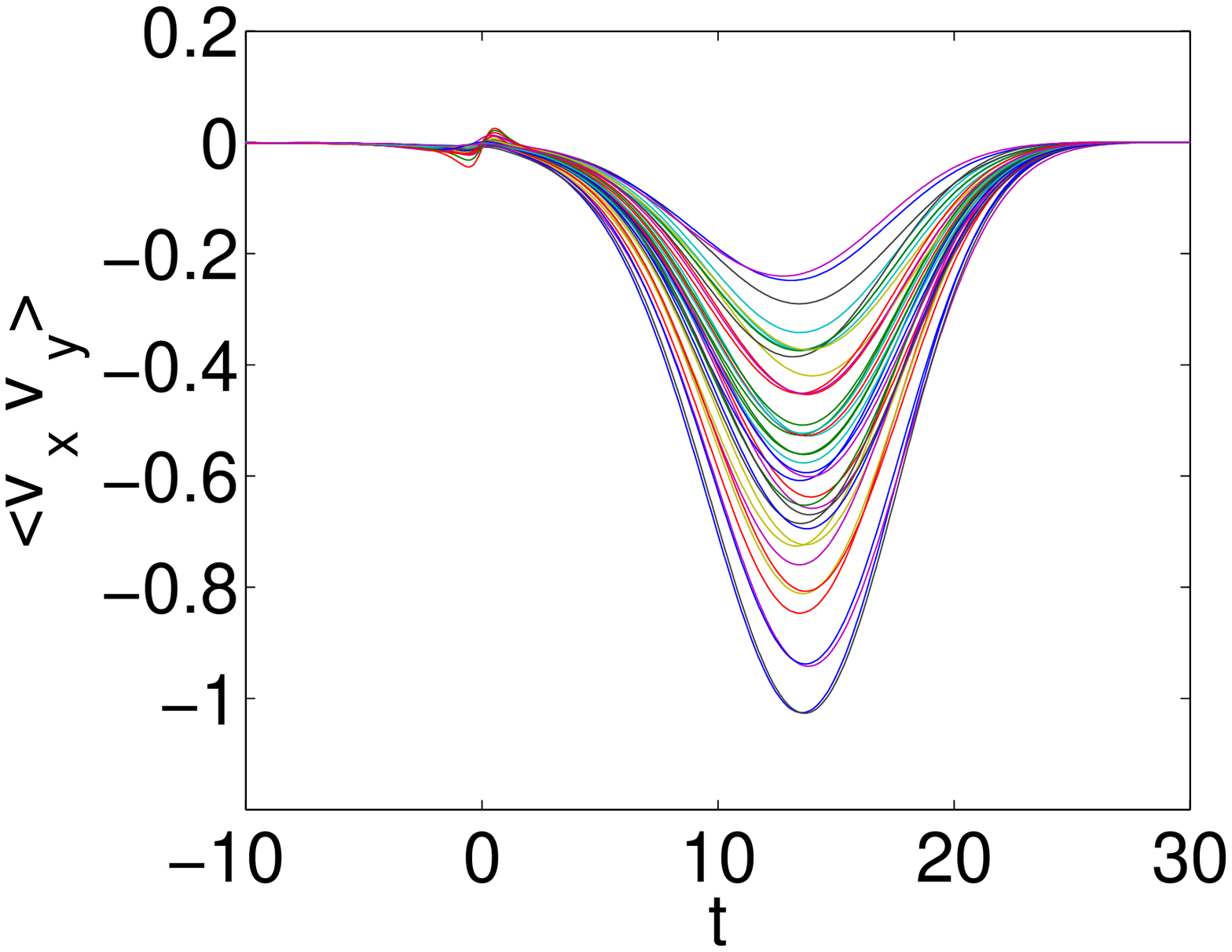}
   \includegraphics[width=0.24\linewidth]{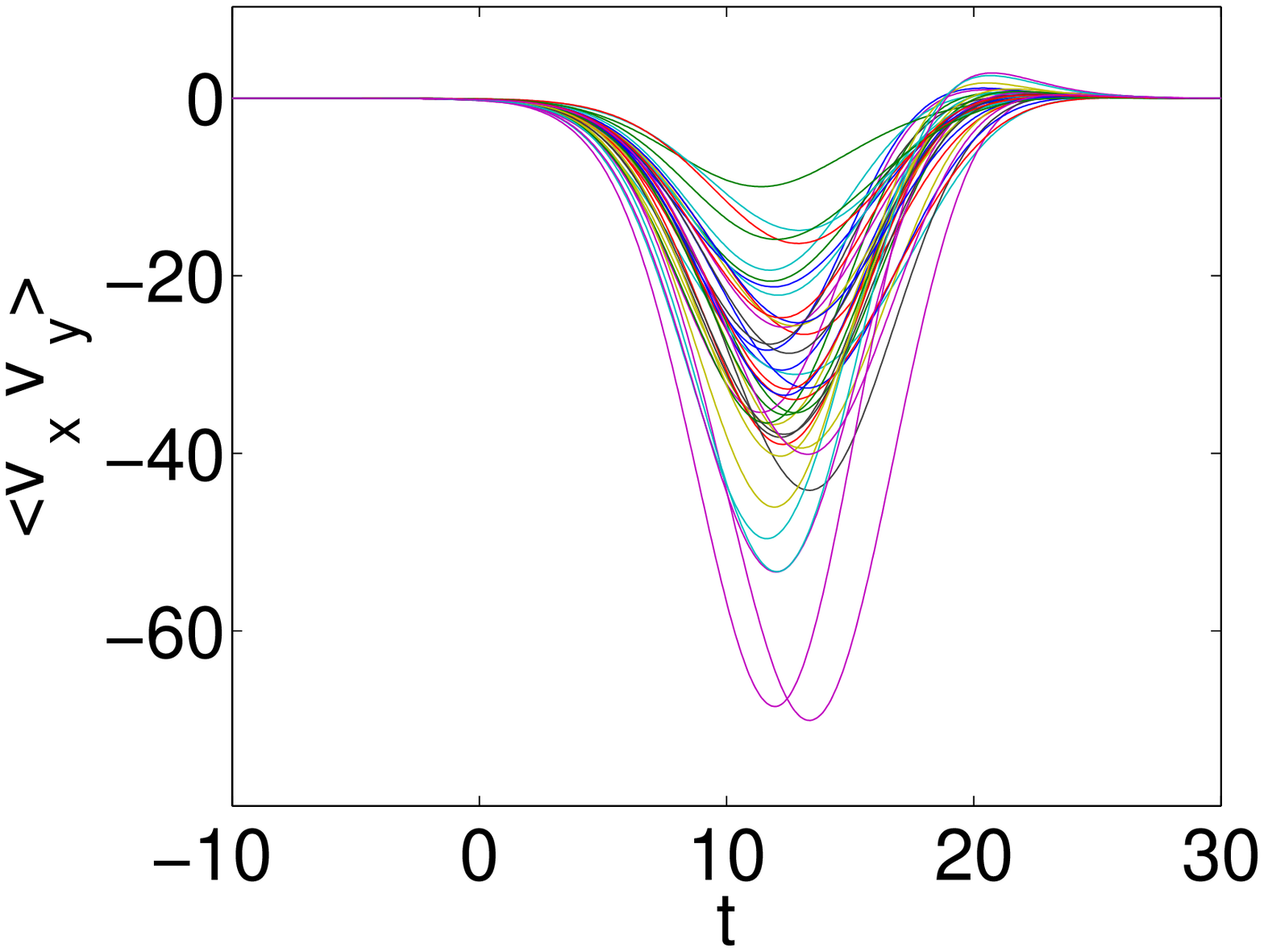}
   \includegraphics[width=0.24\linewidth]{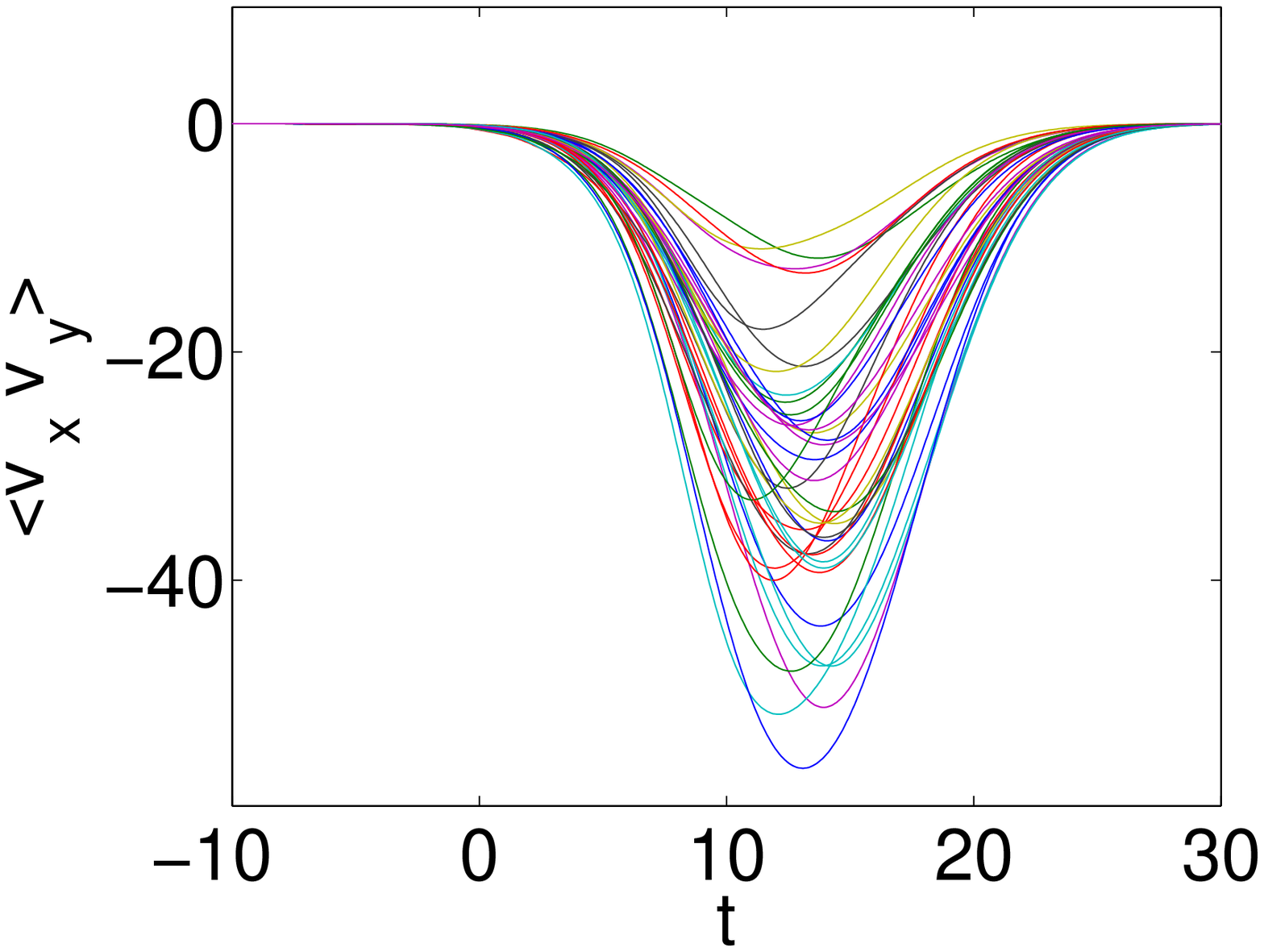}\\
   
   \caption{Evolution of $\langle b_x b_y\rangle$ (upper panels) and $\langle v_x v_y\rangle$ (lower panels) for a set of 40 shearing waves with random initial conditions, $Re=6400$ and $Rm=25600$. From left to right, the large-scale field is increased with $B_0=0.04;0.08;0.2;0.3$. As expected from traditional MRI results, $\langle b_x b_y\rangle>0$ and $\langle v_x v_y\rangle<0$: the shearing waves transport angular momentum outwards.} 
              \label{TransportPlotHRe}%
    \end{figure*}

\subsection{Numerical solution\label{numlin}}
In the following, we will assume a large-scale field similar to the oscillating Fourier mode studied in the previous section. We therefore set:
\begin{equation}
B_x^0(z)=B_0\cos(k_0z).
\end{equation}
The resistive diffusion of this non-uniform field may be neglected on the timescales of interest here. To solve the equations (\ref{motionlin}) and (\ref{inductionlin}), we consider a shearing wave $(k_x,k_y(t))$, $L_z$-periodic in the vertical direction\footnote{The periodicity in the vertical direction is required since the source term $B_x^0(z)$ is periodic with period $L_z$.}. Without any loss of generality, we also define a new time variable $t_\mathrm{SW}$ so that $k_y(t_\mathrm{SW})=0$. To check whether the large-scale field has a systematic impact on the structure of shearing waves, we compute numerically the evolution of several shearing waves using random initial conditions. The shearing waves are initialised at $t_\mathrm{SW}=-10$ with random phases and randomly oriented velocity and magnetic fields. The initial velocity and magnetic fields are verified to satisfy (\ref{Vstruct}) and (\ref{Bstruct}). We then evolve these shearing waves using a Fourier decomposition with $n_z=64$ modes in the vertical direction up to $t=30$. Finally, we compute the axisymmetric EMF due to each shearing wave, $\bm{\mathcal{E}}^\mathrm{SW}(z)=2\Re(\bar{\bm{v}}\times \bar{\bm{b^*}})$, and check the correlation of its curl with the imposed large-scale field, i.e. 
\begin{equation}
\Gamma_j=\varepsilon_{jzm} \frac{\int_0^{L_z} B_x^0(z) \partial_z \mathcal{E}_m^\mathrm{SW}(z)\,dz}{B_0}.
\end{equation}
We plot $\Gamma_x$ and $\Gamma_y$ for a set of 40 shearing waves with $Re=1600$, $Rm=6400$, and the aspect ratio used in the last section, in Fig.~\ref{GammaPlotLRe}. We have considered only the largest nonaxisymmetric waves $k_x=2\pi/L_x$  as they appear to be the dominant ones in the nonlinear simulations (see section \ref{EfieldSection}).
Although the $\Gamma$ coefficients are assumed to be infinitely small compared to $B_0$, they provide an understanding of how a finite-amplitude shearing wave may modify the large-scale field $B_x^0(z)$. To clarify this idea, let us now assume that $B_x^0(z)$ can vary slowly in time. Then, using (\ref{bxbudgeteqn}) and (\ref{bybudgeteqn}), one find that $\Gamma_x>0$ corresponds to a increase of the large-scale field $B_x^0$ whereas $\Gamma_x<0$ is equivalent to a resistive effect. Following the same argument, $\Gamma_y>0$ corresponds to a creation of a large-scale $B_y(z)$ with the same sign as $B_x^0(z)$ which can, in combination with the shear term, amplify $B_x^0(z)$. On the other hand, $\Gamma_y<0$ leads to a $B_y(z)$ with the opposite sign to $B_x^0(z)$ and possibly, in combination with the shear, to a destruction of $B_x^0$. Note however that we have not discussed here the \emph{amplitudes} of the $\Gamma$ coefficients but just their signs. The main reason is that their final amplitude will depend quadratically on the initial excitation of the shearing waves. In a real system, this excitation is a highly nonlinear process which depends on the small-scale properties of the turbulence.\footnote{Note that these properties may depend in turn on the large-scale field and the amplitudes of the shearing waves.} Therefore, our random excitation is not a precise enough turbulence model and we cannot rely on the amplitude found in this linear analysis. Furthermore, the evolution of the shearing waves may differ from the predictions of this analysis if they reach nonlinear amplitudes.

As one can see in Fig.~\ref{GammaPlotLRe}, $\Gamma_x$ has a negative sign on average, independent of $B_0$. As mentioned previously, this can be interpreted as a resistive effect on $B_x^0$ and is similar to the systematic resistive effect observed in numerical simulations (see Fig.~\ref{bxbudget}). This effect is easily explained as resulting from the mixing of the non-uniform azimuthal field by vertical motions. The behaviour of $\Gamma_y$ is a little more complicated since it reverses for $B_0\sim0.08$. For small $B_0$, $\Gamma_y>0$ which implies a possible growth of $B_y(z)$ and therefore an increasing positive shear term in (\ref{bxbudgeteqn}). On the contrary, for $B_0>0.08$, one expects the shear term to decay and eventually be reversed. This property is indeed observed in numerical simulations. In particular, the shear term in Fig.~\ref{bxbudget} starts to decay for $B_0>0.08$ which corresponds to the the predicted result from the linear analysis.

To check the effect of dissipation processes on this picture, we plot in Fig.~\ref{GammaPlotHRe} the evolution of the $\Gamma$ coefficients for Reynolds and magnetic Reynolds numbers 4 times larger than in the previous case. One finds essentially the same properties: a systematic resistive effect for $\Gamma_x$ and a reversal for $\Gamma_y$. Note however that the reversal appears for a larger $B_0$ in the less diffusive case ($B_0\sim 0.2$). We have also checked that these properties persist when one changes the magnetic Prandtl number at sufficiently large $Re$ and $Rm$. These remarks suggest that the results found in this linear analysis are not related to a finite resistivity or viscosity property and may therefore appear for arbitrarily large Reynolds numbers.

\subsection{Phenomenological properties}
The waves described in this section are clearly inhomogeneous in the vertical direction. However, we can understand them as a version of the magnetorotational instability in the presence of a varying azimuthal field \citep{BH92,TP96}. As one would expect, the transport coefficients $\langle b_x b_y\rangle$ and $\langle v_x v_y\rangle$ associated with the linear waves are nonzero, and lead to an outward angular momentum transport (Fig.~\ref{TransportPlotHRe}), as in the classical vertical field calculation \citep[see][]{PCP08}. This demonstrates that these shearing waves actually extract energy from the mean flow as a classical MRI mode. Because of the resistive properties of $\mathcal{E}_y$, these waves also extract some energy from the large-scale azimuthal field, which is generated through shearing of the radial field. 
Therefore, the energy of the waves comes primarily from the mean shear, as expected. 
Moreover, it is known that the azimuthal MRI has an optimum growth rate for a given azimuthal wavelength $k_x$ and field strength $B_x$. In the high-$k_z$ limit with a uniform $\bm{B}=B_0\bm{e_x}$, this maximum growth rate is reached when $k_x B_0=\sqrt{5/12}\,S$ \citep{OP96}. Using our parameters, this leads to $B_0\sim0.2 SL_z$, which is surprisingly close to the magnetic field amplitude for which $\Gamma_x$ is reversed at large $Re$. Although this argument is by no mean comprehensive, it draws attention to the close relation between the transient amplification observed in our analysis and the (ideal) MRI waves studied by \cite{BH92}. Naturally, these conjectures need to be checked more carefully using a proper analytical analysis, which will be the subject of another paper.

\section{A toy model}
In this section, we provide a toy model reproducing the basic linear properties exhibited in the previous section. This toy model does not pretend to be an accurate set of closure relations for equations (\ref{bxbudgeteqn})--(\ref{bybudgeteqn}) but it includes the main physical ingredients required to reproduce qualitatively the cycle behaviour described in section \ref{sectioncycle}. We therefore rewrite equations (\ref{bxbudgeteqn})--(\ref{bybudgeteqn}) as:
\begin{eqnarray}
\label{modelbx}
\partial_t \widehat{B_x}(k_0,t)&=&S\widehat{B_y}(k_0,t)-\beta \widehat{B_x}(k_0,t-t_r),\\
\label{modelby}
\partial_t\widehat{B_y}(k_0,t)&=&\gamma \widehat{B_x}(k_0,t-t_r)\frac{B_r-|\widehat{B_x}(k_0,t-t_r)|}{B_r},
\end{eqnarray}
where $\gamma$, $\beta$ and $B_r$ are constants. In this set of equations, we have neglected the physical dissipation processes, as they are very small in the budget of the simulations (see Figs~\ref{bxbudget}--\ref{bybudget}). We have assumed a resistive term for $\mathcal{E}_y$, as expected from the linear analysis. However, to take into account the fact that the EMF is due to a progressively amplified shearing wave, we assume the EMF is slightly retarded with respect to the large-scale magnetic field, with a delay time $t_r$.  Since the typical shearing wave growth rate is of the order of a shear time, we expect in first approximation $t_r\sim S^{-1}$. The model used for $\mathcal{E}_x$ reproduces the reversal described in the linear analysis, at $|B_x|=B_r$. It also includes the delay used in the $\mathcal{E}_y$ model for the same reasons. This term will be referred to in the following as the $\gamma$ effect term, as it is not formally equivalent to the classical $\alpha$ effect used in dynamo theory, but more an effect of anisotropic turbulent resistivity. Note that the same kind of model involving an anisotropic turbulent resistivity has been studied by \cite{RK03} in the context of the shear dynamo.} For this model to be consistent with the previous analysis, we have to assume that the underlying three-dimensional flow is turbulent in some way, so that small-amplitude shearing waves are continuously excited for $t_\mathrm{SW}<0$, and then amplified linearly. Therefore, this model assumes that the flow is \emph{already} subject to a three-dimensional turbulence and describes the variations of the large-scale field.

\begin{figure}
   \centering
   \includegraphics[width=1.0\linewidth]{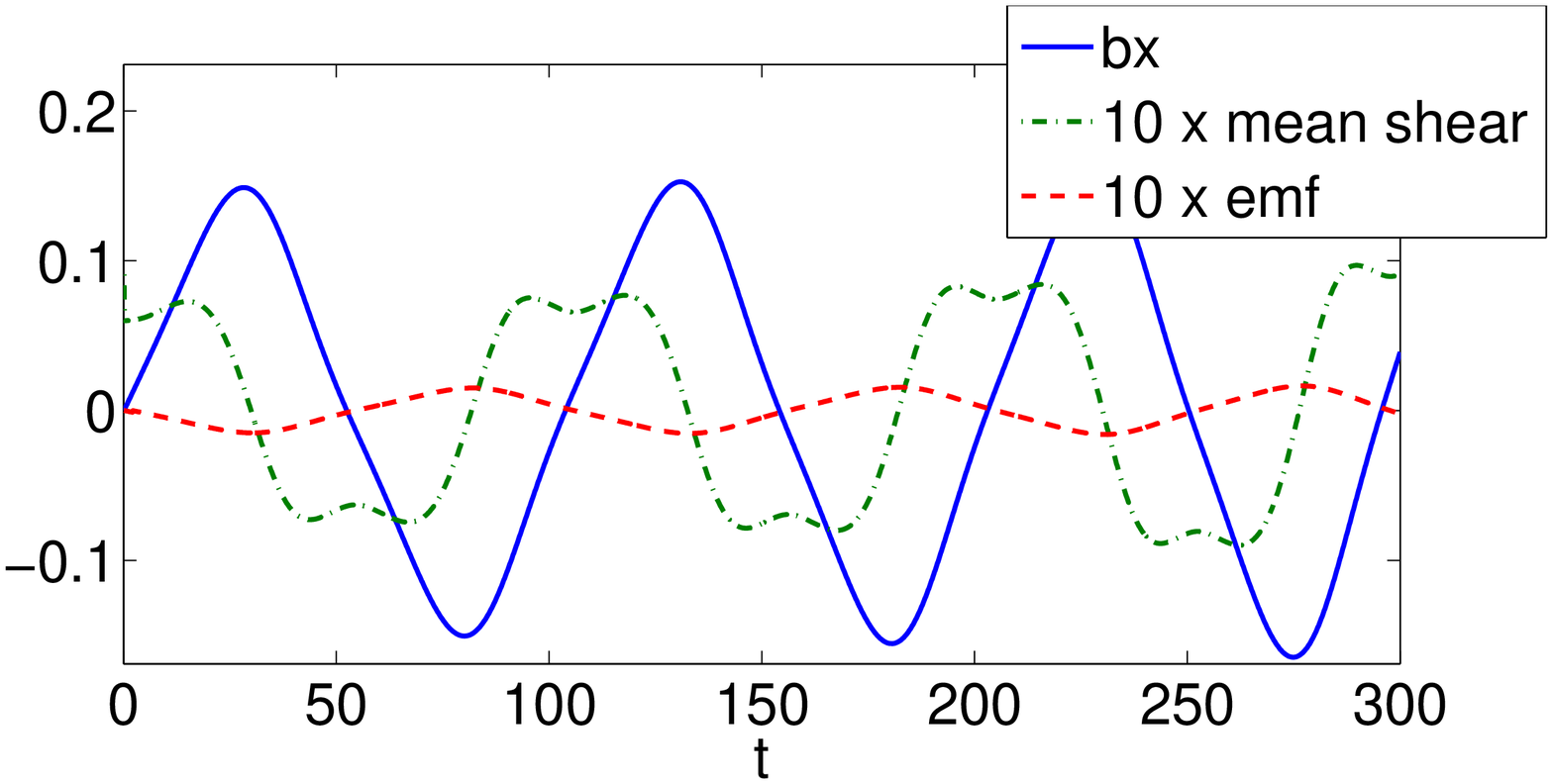}
   \includegraphics[width=1.0\linewidth]{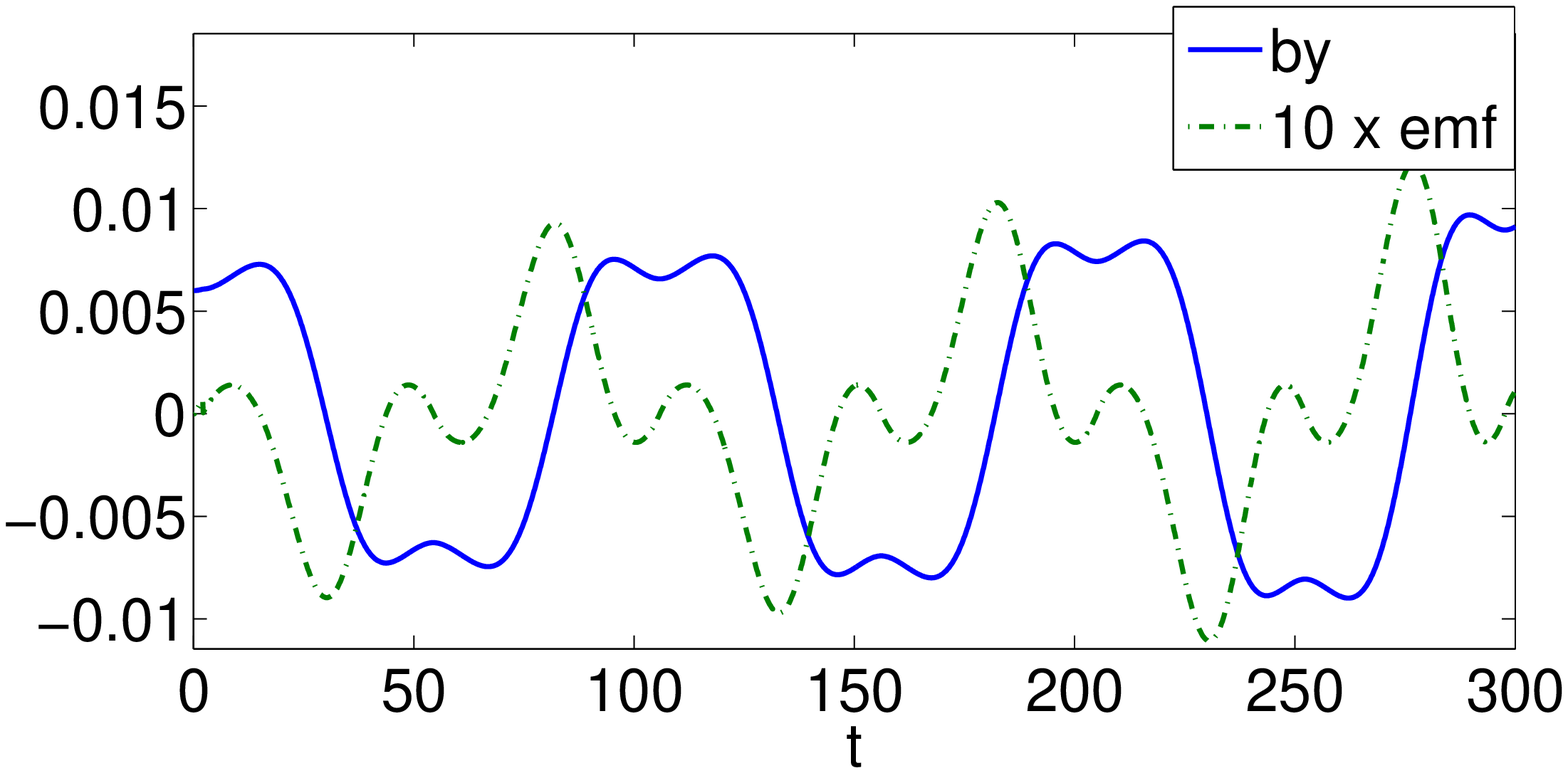}
   \caption{Numerical solution of the toy model (equations \ref{modelbx}--\ref{modelby}) with $\gamma=0.007$, $\beta=0.010$, $t_r=2\,S^{-1}$ and $B_r=0.08$, exhibiting sustained magnetic field oscillations. The time between two $B_x$ reversals  is about $\sim50\,S^{-1}$ as in the full simulations. This toy model reproduces qualitatively the results of our numerical simulations.} 
              \label{BModelPlot}%
    \end{figure}

We numerically solve the toy model using $\gamma=7\times10^{-3}$, $\beta=10^{-2}$, $t_r=2\,S^{-1}$ and $B_r=8\times 10^{-2}$. Here $\gamma$ and $\beta$ are chosen so that our model mimics the main behaviour of the numerical simulations, whereas $B_r$ is chosen according to the results of the linear analysis. The resulting evolution of $B_x$ and $B_y$ is plotted in Fig.~\ref{BModelPlot}, where the ``mean shear'' curve corresponds to the first term on the right-hand side of equation (\ref{modelbx}), and the EMFs are the $\beta$ and $\gamma$ terms. When comparing with the fully nonlinear cycle (Figs~\ref{bxbudget}--\ref{bybudget}), we find essentially the same time history for all quantities. Interestingly, the cycles obtained using the model are self-sustained despite the presence of a resistive effect, showing that the basic properties discussed previously are sufficient to inject magnetic energy into the system. One should note however that the typical EMF term involved in the evolution of $B_x$ is significantly smaller in the model than in the simulation. Moreover, we observe a perfect field reversal of $B_x$ which is clearly not present in the simulations. This may be explained by the presence of ``noise'' due to small turbulent eddies which are continuously exciting the $B_x(k_0)$ mode in the simulations.

As we have seen, our toy model is able to reproduce the basic properties observed in complicated nonlinear simulations, just considering the linear results from the previous section. However, this model is far from perfect. For example, linearizing the model around $\bm{B}=0$ leads to the wrong conclusion that the trivial solution $\bm{B}=0$ is unstable. This peculiar property comes from the assumption of a constant turbulent background. Therefore, if we assume that something (e.g.\ turbulence) is constantly exciting shearing waves, then the solution $\bm{B}=0$ is unstable. However, in a real system, the turbulent motions are partly generated by these shearing waves when they reach a nonlinear regime. If the amplification is not large enough (i.e.\ around $\bm{B}=0$), the shearing waves cannot sustain the turbulence, and therefore the flow cannot excite new shearing waves: the whole system relaxes to $\bm{B}=0$. In conclusion, despite the apparent simplicity of our toy model, the laminar flow is not subject to a linear instability and the underlying mechanism sustaining the turbulence in this system is clearly a nonlinear effect, potentially involving the transient amplification described earlier.

\section{Discussion}
\subsection{Summary}
In this paper, we have investigated the behaviour of the large-scale magnetic field in zero-net-flux simulations of the magnetorotational instability. We have first shown that the large-scale azimuthal field $B_x(z)$ is subject to a long-timescale oscillation when the flow is turbulent. Studying the induction equation, we have seen that these cycles are primarily due to an oscillation of a large-scale radial magnetic field $B_y$ and an azimuthal EMF $\mathcal{E}_x$, whereas the radial EMF $\mathcal{E}_y$ acts like a turbulent resistivity on $B_x$. Studying the nonaxisymmetric modes of the system, we found that the axisymmetric EMF is due mostly to the largest nonaxisymmetric structures, showing that the cycles are essentially a large-scale process. To understand the generation of the EMF, we have investigated the behaviour of nonaxisymmetric linear waves in the presence of an inhomogeneous and constant in time azimuthal magnetic field $B_x(z)$. This linear analysis explains most of the properties of the EMF observed in the nonlinear simulations. In particular, we have shown that the large-scale axisymmetric $\mathcal{E}_x$ can be reversed for large $B_x$, explaining the cycles in the simulations. To summarise these results, we have considered a simplified closure model, encapsulating the linear properties. This model is able to reproduce the main behaviours of the cycles despite its simplicity, showing that the physics involved in the cycles is well described by our linear analysis.

\begin{figure}[h!]
   \centering
   \includegraphics[width=1.0\linewidth]{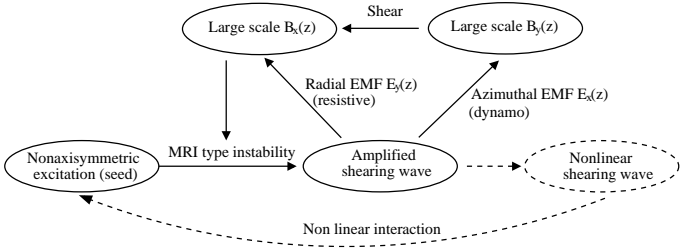}
   \caption{Scheme of the mechanism responsible for the large-scale magnetic field cycles. The dashed lines represent the nonlinear interactions which are not discussed in the present work. These interactions are nonetheless required to obtain a self-sustained mechanism (see text).} 
              \label{mechanism}%
\end{figure}


To summarise our findings, we sketch the mechanism responsible for the cycles in Fig.~\ref{mechanism}. In this sketch, we have assumed that the nonaxisymmetric excitation (seed) is related to the nonlinear coupling of amplified shearing waves (dashed line). This specific process has not been studied in the present work and is just given here as a way to ``close the loop''. Note however that a more complicated mechanism may be involved in this back-reaction without changing the global picture. Moreover, we have described the contribution of $\mathcal{E}_x(z)$ as a generic ``dynamo'' effect, but one should remember that this term may be either correlated or anticorrelated with $B_x$, as a function of the strength of $B_x$. Finally, we have omitted viscous and resistive effects for simplicity, as they have a small impact on the mechanism itself.

\subsection{Comparison with previous works}

We would like to stress that the model presented here does not constitute a full description of a sustaining mechanism for MHD turbulence in discs. Indeed, we have assumed in the linear analysis and in the closure model that the flow is able to generate continuously small-amplitude shearing waves. As mentioned previously, this process involves a nonlinear coupling of nonaxisymmetric waves, which is beyond the scope of this work. However, a fully consistent self-sustaining process would require a description of this effect. According to \cite{FPLH07}, MHD turbulence appears in zero-net-flux shearing-box simulations only when $Pm>1$. However, most of our analysis does not depend on the magnetic Prandtl number, and the dynamo process described here may work for arbitrarily low $Pm$. This suggests that the shearing-wave excitation mechanism depends on the Prandtl number, and may be too weak in the $Pm<1$ case.  This confirms that the excitation should be related to the small turbulent scales, which are still poorly understood.

Interestingly, our results indicate that the large-scale field generation mechanism is not destroyed for large Reynolds numbers, provided that the background flow is turbulent. Therefore, the same kind of mechanism might be at work in real astrophysical discs, generating a large-scale field in a sufficiently ionised and turbulent plasma. One should note however that the configuration used in our simulations is not comparable to the real geometry of an accretion disc. In particular, we do not know how the aspect ratio will modify the mechanism of the cycles, nor how the vertical stratification may enter this picture. However, as discovered by \cite{BNST95} in the case of a stratified flow with nonperiodic vertical boundary conditions, one observes azimuthal field cycles with a long timescale.  These cycles have a significantly longer period ($T\sim 200\, S^{-1}$) compared to ours and involve a modification of the net azimuthal magnetic flux. Moreover, it has been shown by \cite{BD97} that in this case, the cyclic behaviour may be explained by a classical $\alpha$ effect proportional to the vorticity\footnote{Note that this $\alpha$ effect might be related to the presence of vertical stratification and the breaking of the reflectional symmetry}. Because of these fundamental differences, we cannot conclude that the mechanism responsible for all these cycles is the same without further investigation. Nevertheless, it seems that a more systematic investigation of this kind of long-timescale behaviour is required in order to better understand the way MHD turbulence sustains magnetic fields in various configurations.

As mentioned in the introduction, \cite{PCP07} suggested that turbulent transport should vanish for large enough resolution and Reynolds number. Therefore, the turbulent transport associated with this mechanism is another important issue. As one can check easily in the numerical simulation and in the toy model, the large-scale $B_x$ and $B_y$ are roughly phase-shifted by $\pi/2$, leading to a very small Maxwell stress ($\langle B_x B_y\rangle\sim 10^{-4}$). Therefore, in our picture, most of the transport must come from the shearing waves, which have been shown to have nonzero Maxwell and Reynolds stresses. Unfortunately, we cannot put a precise value on the transport due to the shearing waves since it is controlled by their initial excitation. Therefore, this mechanism cannot give a precise answer on the expected transport in a real accretion disc. From a more phenomenological point of view, we expect the large-scale magnetic field strength produced by our mechanism not to depend on the Reynolds numbers if they are sufficiently large (see section \ref{numlin}). Since this large-scale field couples with all the other modes available, we expect the large-scale nonaxisymmetric structures to have roughly the same amplitude as the large-scale field, in the limit of a fully developed turbulence. If this picture is correct, we would then expect a minimum transport of the order of a few times  $10^{-3}$, independently of the Reynolds numbers. Note however that this conclusion relies on the assumption that the large-scale field does not depend on the dissipation coefficients. We have shown that this assumption is plausible, although a precise analytical description of our linear analysis would be required to ascertain this statement.

The present description of an accretion disc dynamo is related to the steady self-sustaining solutions obtained by \cite{ROP07} by numerical continuation methods. Both mechanisms involve a large-scale azimuthal magnetic field with zero net flux, which is generated through the shearing of a radial field.  Both also involve a nonaxisymmetric magnetorotational-type instability that regenerates the radial field through its nonlinear feedback.  However, the present mechanism is intrinsically local and unsteady, and may work for arbitrarily large Reynolds numbers whereas the steady, highly symmetrical solutions of \cite{ROP07} depend on the existence of walls and are apparently restricted to low Reynolds numbers. Nevertheless, one may also think that these steady solutions correspond to some fixed points in phase space, around which the shearing box solutions oscillate, creating the dynamo cycles we have described. If this picture is correct, it would be an elegant way to unify these different approaches, and possibly to find other mechanisms of the same kind. 

This work might also be compared to the shear dynamo simulations of \cite{YH08}. However, we emphasise that our analysis involves a nonlinear dynamo, whereas the shear dynamo described by \cite{YH08} is a kinematic (linear) dynamo, in which turbulence is \emph{forced}. Moreover, the shear dynamo occurs in a non-rotating system in which the MRI does not appear.  Applying our linear analysis to such a system does not produce a $\gamma$ effect that could explain the shear dynamo in either linear or nonlinear regimes.

Finally, we would like to stress that the problem of a non-linear dynamo, involving several linear instabilities, has also been described by \cite{CBC03} in the context of magnetically buoyant flows. Although the underlying instabilities are somewhat different, they also found a cyclic behaviour which, as in the present work, cannot be fully described by a classical $\alpha$ effect. This may indicate that all these instabilities share some common properties yet to be discovered, allowing for the development of a nonlinear dynamo.

\subsection{Future work}
As discussed previously, the main issue raised by our findings is how the turbulence is able to excite shearing waves. Interestingly, the same kind of problem arises in the case of the shearing box with a mean azimuthal field \citep{BH92}. Therefore, a simpler way to study this effect is to investigate the way turbulence is sustained in the presence of a mean azimuthal field. In particular, an interesting test would be to check that in this case, the presence of turbulence also depends on $Pm$, in a similar way as in \cite{FPLH07}. This would be a good indication that the large-scale dynamo process \emph{itself} does not depend on $Pm$ at large $Re$, even though the underlying turbulent regeneration mechanism process does. Moreover, this idea would provide a new way to understand the $Pm$-dependence observed in MRI simulations with a nonzero mean vertical field \citep{LL07}, and potentially to describe the asymptotic transport in the limit of high and low $Pm$.

On the other hand, the dynamo process itself requires further investigation. The linear analysis provided here has been computed using essentially numerical tools, as the analytical description of this problem is somewhat technical. This analytical description is nevertheless required, as it will give a precise and formal constraint on the $\gamma$ effect and its dependence on the azimuthal field strength. It may also be a way to have a more physical understanding of the underlying mechanism responsible for the resistive and $\gamma$ effects, leading to a potential generalisation of this mechanism to a wider class of flow. Finally, we would be able to check, at least linearly, if this dynamo process depends on the non-ideal effects, and potentially to confirm our conjecture.

\begin{acknowledgements}
      We thank John Papaloizou, S\'ebastien Fromang, Pierre-Yves Longaretti and Fran\c{c}ois Rincon for useful discussions. The simulations presented in this paper were performed using the UKMHD cluster based at the University of St Andrews. This research was supported by the Isaac Newton Trust. We thank our anonymous referee for helpful suggestions that significantly improved the paper.
\end{acknowledgements}

\bibliographystyle{aa}
\bibliography{glesur}

\end{document}